\documentclass[aps,floats,twoside,fancyhdr,a4paper,superscriptaddress,nofootinbib,showpacs,preprint]{revtex4}
\usepackage{latexsym}
\usepackage{amsmath}
\usepackage{amssymb}
\usepackage{amsfonts}
\usepackage{graphicx}
\usepackage{dcolumn}
\usepackage{fancyhdr}
\usepackage[colorlinks]{hyperref}

\begin{document}

\title{Full causal bulk viscous LRS Bianchi I with time varying constants}

\author{J. A. Belinch\'on}
\email{abelcal@ciccp.es}
\affiliation{Dpt. of Physics ETS Architecture. UPM Madrid. Av. Juan de Herrera N-4, 28040. Espa\~na.}

\begin{abstract}
In this paper we study the evolution of a LRS Bianchi I Universe,
filled with a bulk viscous cosmological fluid in the presence of time varying constants
 ``but" taking into account the effects of a c-variable into the curvature tensor. We find that the only physical
  models are those which ``constants'' $G$ and $c$ are growing functions on time $t$,
  while the cosmological constant $\Lambda$ is a negative decreasing function.
  In such solutions the energy density obeys the ultrastiff matter equation of state i.e. $\omega=1$.
\end{abstract}

 \vspace{.4cm} \maketitle

\section{Introduction}

In a recent paper (\cite{Tony1})we have studied the behaviour of
the constants $G,c$ and $\Lambda $ within the framework of a flat
FRW cosmological model where we consider the effects of a
c-variable into the curvature tensor. The study was restricted to
the perfect fluid case. In such paper we arrived to the conclusion
that we have only two physical solutions and furthermore both
solutions follow a power-law time dependence of the physical
quantities on the cosmological time
 while in the previous works, where we have not taken into
account the effects of a c-variable into the curvature tensor, we
were capable of obtaining more solutions. We would like to
emphasize that in such study we obtained as integration condition
that both ``constants'', in spite of considering that they vary,
verify the relationship $G/c^{2}=const.$ for all equation of
state. Nevertheless we have not been able to find any physical
restriction in the parameters to determine if the ``constants''
are growing or decreasing functions on time $t,$ since both cases
are permissible. We would like to point out this fact, since
``traditionally'' until now all the solutions obtained for both
``constants'' (when they vary simultaneously) are decreasing on
time $t.$ In such works it was not considered the possible effects
of a c-variable into the curvature tensor in such a way that the
traditional FRW equations remain unalterable.

In the present paper we try to generalize the study begun in
(\cite{Tony1}) considering a cosmological model with symmetries
LRS Bianchi I with variable constants and which momentum-energy
tensor is modeled by a bulk viscous fluid (full causal theory) in
such a way and expecting that the thermodynamics help us to
restrict the parameters to find physical solutions. Furthermore
bulk viscosity is expected to play an important role in the early
evolution of the Universe, when also the dynamics of the
``constants'' $G,c$ and $\Lambda $ could be different.
 We shall consider the
possible effects of a c-variable into the curvature tensor in such
a way that we shall outlined a new field equations and we shall
study the Krestchmann invariants as well as the expansion and the
shear. Once the field equations have been outlined we will need to
make simplifying hypotheses in order to try to obtain a complete
solution to the field equations.

Under the assumed hypotheses we will study three different models
finding that the physically reasonable models (decreasing
temperature, increasing entropy, negative viscous pressure etc...)
are those with $G$ and $c$ as growing function on time $t$ and in
particular such models have an equation of the state $\omega =1$
that is to say, ultrastiff matter, and the viscous parameter is
$\gamma =1/2$, the usual one. In such solutions the cosmological
``constant'' is a negative decreasing function. The models which
have decreasing functions $G$ and $c$ result without any physical
sense i.e. positive viscous pressure, decreasing entropy etc....
We end the paper summarizing all these results.

\section{The model and the hypotheses}

A locally rotationally symmetric (LRS) Bianchi I (\cite{Ha, Prad}) spacetime with
metric
\begin{equation}
ds^{2}=-c^{2}(t)dt^{2}+X^{2}(t)dx^{2}+Y^{2}(t)\left( dy^{2}+dz^{2}\right) ,
\label{line}
\end{equation}
filled with a bulk viscous cosmological fluid with the following
energy-momentum tensor \cite{Ma95, Ma96}:
\begin{equation}
T_{i}^{k}=\left( \rho +p+\Pi \right) u_{i}u^{k}-\left( p+\Pi \right) \delta
_{i}^{k},  \label{1}
\end{equation}
where $\rho $ is the energy density, $p$ the thermodynamic pressure, $\Pi $
is the bulk viscous pressure and $u_{i}$ is the four velocity satisfying the
condition $u_{i}u^{i}=-1$. The number $4-$flux and the entropy $4-$flux take
the form:
\begin{align}
N^{i}& =nu^{i},  \label{num} \\
S^{i}& =sN^{i}-\left( \frac{\tau \Pi ^{2}}{2\xi T}\right) u^{i},
\label{entropy}
\end{align}
where $n$ is the number density, $s$ the specific entropy, $T\geq 0$ the
temperature, $\xi $ the bulk viscosity coefficient and $\tau \geq 0$ the
relaxation coefficient for transient bulk viscous effect (i.e. the
relaxation time).

The fundamental thermodynamic tensors (\ref{1}-\ref{entropy}) are subject to
the dynamical laws of energy-momentum conservation, number conservation and
the Gibb's equation:
\begin{align}
T_{i;k}^{k} & =0,  \label{con1} \\
N_{\text{ };i}^{i} & =0,  \label{con2} \\
Tds & =d\left( \frac{\rho}{n}\right) +pd\left( \frac{1}{n}\right) .
\label{con3}
\end{align}

The equations (\ref{1}-\ref{entropy}) and (\ref{con1}-\ref{con3}) imply:
\begin{equation}
TS_{;i}^{i}=-\Pi \left( 3H+\frac{\dot{\tau}}{\xi }\dot{\Pi}+\frac{1}{2}T\Pi
\left( \frac{\tau }{\xi T}u^{i}\right) _{;i}\right) ,  \label{preequation}
\end{equation}
(where $3H=u_{;i}^{i})$ by equations (\ref{con1}-\ref{con3}) and (\ref
{preequation}), the simplest way (linear in $\Pi )$ to satisfy the $H$%
-theorem (i.e. for the entropy production to be non-negative, $S_{;i}^{i}=%
\frac{\Pi ^{2}}{\xi T}\geq 0$ $)$ leads to the causal evolution equation for
bulk viscosity given by \cite{Ma95}
\begin{equation}
\tau \dot{\Pi}+\Pi =-3\xi H-\frac{\epsilon }{2}\tau \Pi \left( 3H+\frac{\dot{%
\tau}}{\tau }-\frac{\dot{\xi}}{\xi }-\frac{\dot{T}}{T}\right) ,  \label{bulk}
\end{equation}

In eq.(\ref{bulk}), $\epsilon=0$ gives \ the truncated theory (the truncated
theory implies a drastic condition on the temperature), while $\epsilon=1$
gives the full theory. The non-causal theory has $\tau=0$.

The growth of the total commoving entropy $\Sigma =snV$ over a proper time
interval $\left( t_{0},t\right) $ is given by \cite{Ma95}:
\begin{equation}
\Sigma (t)-\Sigma \left( t_{0}\right) =-\frac{3}{k_{B}}\int_{t_{0}}^{t}\frac{%
\Pi HV}{T}dt,  \label{M entropy}
\end{equation}
where $k_{B}$ is the Boltzmann's constant, and $V=\left( XY^{2}\right) $ is
the proper volume.

The Einstein gravitational field equations with variable $G$, $c$ and $%
\Lambda $ are:
\begin{equation}
R_{ik}-\frac{1}{2}g_{ik}R=\frac{8\pi G(t)}{c^{4}\left( t\right) }%
T_{ik}+\Lambda (t)g_{ik}.  \label{ECU1}
\end{equation}
Applying the covariance divergence to the second member of equation (\ref
{ECU1}) we get:
\begin{equation}
div\left( \frac{G}{c^{4}}T_{i}^{j}+\delta _{i}^{j}\Lambda \right) =0,
\label{conser1}
\end{equation}
\begin{equation}
T_{i;j}^{j}=\left( \frac{4c_{,j}}{c}-\frac{G_{,j}}{G}\right) T_{i}^{j}-\frac{%
c^{4}\delta _{i}^{j}\Lambda _{,j}}{8\pi G},  \label{conser2}
\end{equation}
that simplifies to:
\begin{equation}
\dot{\rho}+3\left( \rho +p\right) H+3H\Pi =-\frac{\overset{\cdot }{\Lambda }%
c^{4}}{8\pi G}-\rho \frac{\dot{G}}{G}-4\rho \frac{\dot{c}}{c},
\label{conser3}
\end{equation}

Therefore, our model (with LRS BI symmetries) is described by the following
equations:
\begin{align}
2\frac{\dot{X}}{X}\frac{\dot{Y}}{Y}+\frac{\dot{Y}^{2}}{Y^{2}}& =\frac{8\pi G%
}{c^{2}}\rho +\Lambda c^{2},  \label{field1} \\
2\frac{\ddot{Y}}{Y}+\frac{\dot{Y}^{2}}{Y^{2}}-2\frac{\dot{Y}}{Y}\frac{\dot{c}%
}{c}& =-\frac{8\pi G}{c^{2}}\left( p+\Pi \right) +\Lambda c^{2},
\label{field2} \\
\frac{\ddot{Y}}{Y}+\frac{\dot{X}}{X}\frac{\dot{Y}}{Y}+\frac{\ddot{X}}{X}%
-\left( \frac{\dot{X}}{X}+\frac{\dot{Y}}{Y}\right) \frac{\dot{c}}{c}& =-%
\frac{8\pi G}{c^{2}}\left( p+\Pi \right) +\Lambda c^{2},  \label{field2-2} \\
\dot{\rho}+\left( \rho +p+\Pi \right) \left( \frac{\dot{X}}{X}+2\frac{\dot{Y}%
}{Y}\right) & =-\frac{\dot{\Lambda}c^{4}}{8\pi G}-\rho \frac{\dot{G}}{G}%
+4\rho \frac{\dot{c}}{c},  \label{field3} \\
\tau \dot{\Pi}+\Pi +\xi \left( \frac{\dot{X}}{X}+2\frac{\dot{Y}}{Y}\right) &
=-\frac{\epsilon }{2}\tau \Pi \left( \left( \frac{\dot{X}}{X}+2\frac{\dot{Y}%
}{Y}\right) +\frac{\dot{\tau}}{\tau }-\frac{\dot{\xi}}{\xi }-\frac{\dot{T}}{T%
}\right) .  \label{field4}
\end{align}

In order to close the system of equations (\ref{field1}-\ref{field4}) we
have to give the equation of state for $p$ and specify $T$, $\xi$ and $\tau$%
. As usual, we assume the following phenomenological (ad hoc) laws \cite
{Ma95}:
\begin{align}
p & =\omega\rho,  \label{csi1} \\
\xi & =k_{\gamma}\rho^{\gamma},  \label{csi2} \\
T & =D_{\delta}\rho^{\delta},  \label{csi3} \\
\tau & =\xi\rho^{-1}=k_{\gamma}\rho^{\gamma-1},  \label{csi4}
\end{align}
where $0\leq\omega\leq1$, and $k_{\gamma}\geq0$, $D_{\delta}\geq0$ are
dimensional constants, $\gamma\geq0$ and $\delta\geq0$ are numerical
constants. Eqs. (\ref{csi1}) are standard in cosmological models whereas the
equation for $\tau$ is a simple procedure to ensure that the speed of
viscous pulses does not exceed the speed of light. These are without
sufficient thermodynamical motivation, but, in absence of better
alternatives, we use these equations and expect that they will at least
provide an indication of the range of possibilities. For the temperature
law, we take, $T=D_{\delta }\rho^{\delta}$, which is the simplest law
guaranteeing positive heat capacity.

In the context of irreversible thermodynamics $p$, $\rho$, $T$ and the
particle number density $n$ are equilibrium magnitudes which are related by
equations of state of the form $\rho=\rho(T,n)$ and $p=p(T,n)$. From the
requirement that the entropy is a state function, we obtain the equation

\begin{equation}
\left( \frac{\partial \rho }{\partial n}\right) _{T}=\frac{p+\rho }{n}-\frac{%
T}{n}\left( \frac{\partial p}{\partial T}\right) _{n}.  \label{thermo1}
\end{equation}
For the equations of state (\ref{csi1}-\ref{csi4}) this relation imposes the
constraint $\delta =\frac{\omega }{\omega +1}$ so that $0\leq \delta \leq
1/2 $ for $0\leq \omega \leq 1$, a range of values which is usually
considered in the physical literature .

The Israel-Stewart-Hiscock theory is derived under the assumption that the
thermodynamical state of the fluid is close to equilibrium, that is, the
non-equilibrium bulk viscous pressure should be small when compared to the
local equilibrium pressure $\left| \Pi \right| <<p=\omega \rho $ . If this
condition is violated then one is effectively assuming that the linear
theory holds also in the nonlinear regime far from equilibrium. For a fluid
description of the matter, this condition should be satisfied.

Therefore, with all these assumptions and taking into account the
conservation principle, i.e., $div(T_{i}^{j})=0$, the resulting field
equations are as follows:

\begin{align}
2\frac{\dot{X}}{X}\frac{\dot{Y}}{Y}+\frac{\dot{Y}^{2}}{Y^{2}}& =\frac{8\pi G%
}{c^{2}}\rho +\Lambda c^{2},  \label{nfield1} \\
2\frac{\ddot{Y}}{Y}+\frac{\dot{Y}^{2}}{Y^{2}}-2\frac{\dot{Y}}{Y}\frac{\dot{c}%
}{c}& =-\frac{8\pi G}{c^{2}}\left( p+\Pi \right) +\Lambda c^{2},
\label{nfield2} \\
\frac{\ddot{Y}}{Y}+\frac{\dot{X}}{X}\frac{\dot{Y}}{Y}+\frac{\ddot{X}}{X}%
-\left( \frac{\dot{X}}{X}+\frac{\dot{Y}}{Y}\right) \frac{\dot{c}}{c}& =-%
\frac{8\pi G}{c^{2}}\left( p+\Pi \right) +\Lambda c^{2},  \label{nfield22} \\
\dot{\rho}+\left( \omega +1\right) \rho \left( \frac{\dot{X}}{X}+2\frac{\dot{%
Y}}{Y}\right) & =-\left( \frac{\dot{X}}{X}+2\frac{\dot{Y}}{Y}\right) \Pi ,
\label{nfield3} \\
\frac{\dot{\Lambda}c^{4}}{8\pi G}+\rho \frac{\dot{G}}{G}-4\rho \frac{\dot{c}%
}{c}& =0,  \label{nfield4} \\
\dot{\Pi}+\frac{\Pi }{k_{\gamma }\rho ^{\gamma -1}}+\rho \left( \frac{\dot{X}%
}{X}+2\frac{\dot{Y}}{Y}\right) & =-\frac{1}{2}\Pi \left( \left( \frac{\dot{X}%
}{X}+2\frac{\dot{Y}}{Y}\right) -W\frac{\dot{\rho}}{\rho }\right) ,
\label{nfield5}
\end{align}
where $W=\left( \frac{2\omega +1}{\omega +1}\right) $.

Since we have defined the $4-$velocity as:
\begin{equation}
u^{i}=\left( \frac{1}{c(t)},0,0,0\right) \text{ \ \ \ \ \ \ \ / \ \ \ \ \ \
\ }u^{i}u_{j}=-1
\end{equation}
then the expansion is defined as follows
\begin{equation}
\theta :=u_{\,;\,i}^{i}\text{, \ \ \ \ \ \ \ \ \ }\theta =\frac{1}{c(t)}%
\left( \frac{\dot{X}}{X}+2\frac{\dot{Y}}{Y}\right)  \label{expan}
\end{equation}
and the shear is
\begin{equation}
\sigma ^{2}=\frac{1}{2}\sigma _{ij}\sigma ^{ij},\text{ \ \ \ \ \ \ }\sigma =%
\frac{\sqrt{3}}{3c(t)}\left( \frac{\dot{X}}{X}-\frac{\dot{Y}}{Y}\right)
\label{shear}
\end{equation}

Curvature is described by the tensor field $R_{jkl}^{i}.$ It is well know
that if one uses the singular behaviour of the tensor components \ or its
derivates as a criterion for singularities, one gets into trouble since the
singular behaviour of the coordinates or the tetrad basis rather than the
curvature tensor. To avoid this problem, one should examine the scalars
formed out of the curvature. The invariants $K_{1}$ and $K_{2}$ (the
Kretschmann scalars) are very useful for the study of the singular
behaviour:
\begin{eqnarray}
K_{1} &:&=R_{ijkl}R^{ijkl}=\frac{4}{c^{4}}\left[ \left( \frac{\ddot{X}}{X}%
\right) ^{2}-2\frac{\ddot{X}}{X}^{2}\frac{\dot{c}\dot{X}}{cX}+\frac{\dot{c}%
^{2}\dot{X}^{2}}{c^{2}X^{2}}+2\left( \frac{\ddot{Y}}{Y}\right) ^{2}-\right.
\\
&&\left. -4\frac{\ddot{Y}}{Y}\frac{\dot{Y}}{Y}\frac{\dot{c}}{c}+\left( \frac{%
\dot{Y}\dot{c}}{Yc}\right) ^{2}+2\left( \frac{\dot{X}}{X}\frac{\dot{Y}}{Y}%
\right) ^{2}+\left( \frac{\dot{Y}}{Y}\right) ^{4}\right]  \label{k1}
\end{eqnarray}
\begin{eqnarray}
K_{2} &:&=R_{ij}R^{ij}=\frac{2}{c^{4}}\left[ \left( \frac{\ddot{X}}{X}%
\right) ^{2}+\left( \frac{\dot{c}\dot{X}}{cX}\right) ^{2}-2\frac{\ddot{X}}{X}%
^{2}\frac{\dot{c}\dot{X}}{cX}+3\left( \frac{\dot{X}}{X}\frac{\dot{Y}}{Y}%
\right) ^{2}+\left( \frac{\dot{Y}}{Y}\right) ^{4}+3\left( \frac{\ddot{Y}}{Y}%
\right) ^{2}\right. \\
&&+3\left( \frac{\dot{Y}}{Y}\frac{\dot{c}}{c}\right) ^{2}-6\frac{\ddot{Y}}{Y}%
\frac{\dot{Y}}{Y}\frac{\dot{c}}{c}+2\frac{\ddot{X}}{X}\frac{\ddot{Y}}{Y}-2%
\frac{\ddot{X}}{X}\frac{\dot{Y}}{Y}\frac{\dot{c}}{c}-2\frac{\ddot{Y}}{Y}%
\frac{\dot{c}\dot{X}}{cX}+2\frac{\dot{X}}{X}\frac{\dot{Y}}{Y}\left( \frac{%
\dot{c}}{c}\right) ^{2}+ \\
&&+2\frac{\dot{X}}{X}\frac{\dot{Y}}{Y}\frac{\ddot{X}}{X}-2\left( \frac{\dot{X%
}}{X}\right) ^{2}\frac{\dot{Y}}{Y}\frac{\dot{c}}{c}+2\frac{\ddot{Y}}{Y}%
\left( \frac{\dot{Y}}{Y}\right) ^{2}+2\frac{\ddot{Y}}{Y}\frac{\dot{Y}}{Y}%
\frac{\dot{X}}{X}-2\frac{\dot{X}}{X}\left( \frac{\dot{Y}}{Y}\right) ^{2}%
\frac{\dot{c}}{c}- \\
&&\left. -2\left( \frac{\dot{Y}}{Y}\right) ^{3}\frac{\dot{c}}{c}+2\left(
\frac{\dot{Y}}{Y}\right) ^{3}\frac{\dot{X}}{X}\right]  \label{K2}
\end{eqnarray}

\subsection{Simplifying hypotheses}

In order to solve these equations we will need to make the following
simplifying hypotheses:

\begin{enumerate}
\item  We assume a relationship between the spatial directions
\begin{equation}
X\thickapprox Y^{n}
\end{equation}
with $n\neq 1,$ and we will impose that the bulk viscous pressure
follows a similar behavior as the energy density i.e.
\begin{equation}
\Pi \thickapprox \rho,  \label{H1}
\end{equation}
since $\left[ \Pi \right] =\left[ \rho \right] .$

The bulk viscous pressure and the energy density of the cosmological fluid
are proportional to each other and hence, the general evolution of $\Pi$ is
qualitatively similar to the evolution of the thermodynamic pressure $p$,
both obeying a similar equation of state. In fact, by defining an effective
coefficient $\omega_{eff}=\omega-\varkappa$, the equation of state of the
cosmological fluid can be formally written as $p_{eff}=p+\Pi=\left(
\omega_{eff}-1\right) \rho$. However, since the bulk viscous pressure of the
cosmological fluid must also satisfy the evolution equation (\ref{nfield5}),
the resulting time evolution depends not only on $\omega_{eff}$, but also,
via the coefficients $\delta$ and $\gamma$, on the equations of state of the
bulk viscosity coefficient, $\xi=\xi\left( \rho\right) $ and of the
temperature, $T=T\left( \rho\right) $. Therefore, even that formally one can
introduce an effective pressure (including both $p$ and $\Pi$), obeying a $%
\omega$-law equation of state, due to the extra-constraints imposed by the
requirements of the causal bulk evolution, the general dynamics of the
present model cannot be reduced to a perfect fluid model evolution.

\item  As in previous works  \cite{Ha, M}, but here, with the difference of considering a c-variable, we see from the equations (\ref{nfield2}) and (\ref{nfield22}) that  it is
obtained
\begin{equation}
\frac{\ddot{Y}}{Y}+\frac{\dot{Y}^{2}}{Y^{2}}+\left( \frac{\dot{X}}{X}-\frac{%
\dot{Y}}{Y}\right) \frac{\dot{c}}{c}-\frac{\dot{X}}{X}\frac{\dot{Y}}{Y}-%
\frac{\ddot{X}}{X}=0  \label{M_T}
\end{equation}

\begin{enumerate}
\item  then we can make the next hypothesis: we impose that (as in the above
case)
\begin{equation}
X\thickapprox Y^{n},
\end{equation}
therefore equation (\ref{M_T}) yields
\begin{equation}
\ddot{Y}=\left( \frac{n^{2}-1}{1-n}\right) \frac{\dot{Y}^{2}}{Y}+\dot{Y}%
\frac{\dot{c}}{c}
\end{equation}
obtaining a trivial solution
\begin{equation}
Y=\left( C_{2}(n+2)+C_{1}(n+2)\int c(t)dt\right) ^{\frac{1}{(n+2)}}
\end{equation}
then we will need to make a hypothesis about the behaviour of $c(t)$ in
order to obtain a complete solution.

\item  if we make the next assumption
\begin{equation}
c(t)=c_{0}Y^{n}
\end{equation}
then eq. (\ref{M_T}) has the following solution:
\begin{equation}
Y=X\left( C_{2}\left( n-2\right) \int X^{n-3}dt+C_{1}(2-n)\right) ^{\frac{1}{%
2-n}}
\end{equation}
then we will need to make a hypothesis about the behaviour of $X(t)$ in
order to obtain a complete solution.
\end{enumerate}
\end{enumerate}

In the subsequent sections we shall study three different models
which correspond to each one of the hypotheses.

\section{Solutions}

\subsection{Model 1. Hypotheses 1.}

We formalize this relationship in the following
way, $\Pi =\varkappa \rho $, and under physical considerations we take $%
\varkappa \in \mathbb{R}^{-}$ and $\left| \varkappa \right| \ll 1$ \cite
{Zi96, Zi04}$.$

In this section, we integrate the field equations with these assumptions and
taking into account the conservation principle. This leads us to the
following relationship, since
\begin{equation}
X\thickapprox Y^{n}\Longrightarrow \dot{\rho}+\left( \omega +1+\varkappa
\right) \rho (n+2)\left( \frac{\dot{Y}}{Y}\right) =0.
\end{equation}
This trivially lead us to the well known relationship between the energy
density $\rho $ and the scale factor $Y$
\begin{equation}
\rho =A_{\omega }Y^{-\left( \omega +1+\varkappa \right) (n+2)}\text{ \ \ \
or \ \ \ \ \ }\rho =A_{\omega }Y^{-\alpha },  \label{INT conservation}
\end{equation}
where $\alpha =\left( \omega +1+\varkappa \right) (n+2).$ It is observed
that if $n=1$, we recover the standard FRW model.

Now, taking into account the eq. (\ref{nfield5}) \ and simplifying it, we
obtain,
\begin{equation}
\varkappa \dot{\rho}+\frac{\varkappa \rho }{k_{\gamma }\rho ^{\gamma -1}}=%
\frac{\left( n+2\right) }{\alpha }\dot{\rho}-\frac{1}{2}\varkappa \dot{\rho}%
\left( -\frac{\left( n+2\right) }{\alpha }-W\right) ,
\end{equation}
thereby obtaining $\rho =\rho (t).$ On simplifying further, we obtain,
\begin{equation}
\frac{\dot{\rho}}{\rho ^{2-\gamma }}=\frac{K}{k_{\gamma }}\text{ \ \ \ }%
\Longrightarrow \text{ \ \ \ }\rho =dk_{\gamma }^{-b}t^{b},
\label{INT density}
\end{equation}
where
\begin{equation}
\text{\ \ \ }K=\left( \frac{\varkappa }{\frac{\left( n+2\right) }{\alpha }+%
\frac{\left( n+2\right) \varkappa }{2\alpha }+\frac{W\varkappa }{2}%
-\varkappa }\right) ,\text{ \ }d=\left( \gamma -1\right) K^{b}\text{\ and \
\ \ \ }b=\frac{1}{\gamma -1}.
\end{equation}

From equation\ (\ref{INT conservation}) we obtain:
\begin{equation}
Y=\left( \frac{A_{\omega }}{d}k_{\gamma }^{b}t^{-b}\right) ^{1/\alpha },%
\text{ \ \ i.e. \ \ }Y\propto t^{\frac{-1}{\left( \omega +1+\varkappa
\right) \left( \gamma -1\right) \left( n+2\right) }}.  \label{INT sfactor}
\end{equation}

An important observational quantity is the deceleration parameter $q=\frac{d%
}{dt}\left( \frac{1}{H}\right) -1$, where $H=\frac{1}{3}\sum_{i}\frac{\dot{a}%
_{i}}{a_{i}}=\frac{1}{3}\left( \frac{\dot{X}}{X}+2\frac{\dot{Y}}{Y}\right) ,$
in this case $H=\frac{n+2}{3}\frac{\dot{Y}}{Y}.$ The sign of the
deceleration parameter indicates whether the model inflates or not. The
positive sign of $q$ corresponds to ``standard'' decelerating models whereas
the negative sign indicates inflation. In our model, the deceleration
parameter behaves as:
\begin{equation}
q=-1-\frac{3\alpha }{b(n+2)}.
\end{equation}
Thus, for the parameters $\gamma ,\omega $ and $\varkappa $, the
deceleration parameter indicates an inflationary behaviour when 3$\alpha
/b(n+2)>0 $ and a non-inflationary behaviour for $\alpha /b(n+2)<-1$.

We proceed with the calculation of the other physical quantities as under:
\begin{align}
\xi & =k_{\gamma }\rho ^{\gamma }\propto k_{\gamma }\left( dk_{\gamma
}^{-b}t^{b}\right) ^{\gamma }=d^{\gamma }k_{\gamma }^{1-b\gamma }t^{\gamma
b}, \\
T& =D_{\delta }\rho ^{\delta }=D_{\delta }\left( dk_{\gamma
}^{-b}t^{b}\right) ^{\delta }\text{ \ \ \ with \ \ }\delta =\frac{\omega }{%
\omega +1}, \\
\tau & =\xi \rho ^{-1}=k_{\gamma }\left( dk_{\gamma }^{-b}t^{b}\right)
^{\left( \gamma -1\right) },\text{ i.e. }\tau =d^{\gamma -1}t.
\end{align}
We see from $\tau =d^{\gamma -1}t$ that this result is in agreement with the
theoretical result obtained in \cite{Ma95}. For viscous expansion to be
non-thermalizing, we should have $\tau <t,$ or otherwise the basic
interaction rate for viscous effects should be sufficiently rapid to restore
the equilibrium as the fluid expands.

The comoving entropy is taking into account the relationship
\begin{equation}
\Sigma (t)-\Sigma \left( t_{0}\right) =\frac{\varkappa (n+2)}{d^{\frac{(n+2)b%
}{\alpha }+\delta -1}}\frac{A_{\omega }^{(n+2)/\alpha }}{\alpha
k_{B}D_{\delta }}\frac{k_{\gamma }^{b(\delta +\frac{n+2}{\alpha }-1)}}{(1-%
\frac{n+2}{\alpha }-\delta )}\left[ \tilde{t}^{b(1-\frac{n+2}{\alpha }%
-\delta )}\right] _{t_{0}}^{t},
\end{equation}
\begin{equation}
\Sigma (t)-\Sigma \left( t_{0}\right) =\Sigma _{0}\left[ \tilde{t}^{b(1-%
\frac{n+2}{\alpha }-\delta )}\right] _{t_{0}}^{t}
\end{equation}
We notice that the parameter $\varkappa $ weakly perturbs the perfect fluid
FRW Universe. When $\varkappa =0$, the comoving entropy assumes a constant
value and we recover the perfect fluid case. Finally, we will use the
equations
\begin{align}
\left( 2n+1\right) \frac{\dot{Y}^{2}}{Y^{2}}& =\frac{8\pi G}{c^{2}}\rho
+\Lambda c^{2},  \label{sil1} \\
\frac{\dot{\Lambda}c^{4}}{8\pi G\rho }+\frac{\dot{G}}{G}-4\frac{\dot{c}}{c}&
=0,  \label{sil2}
\end{align}
to obtain the behaviour of the ``constants'' $G,c$ and $\Lambda .$ From (\ref
{sil1}), we obtain $\dot{\Lambda}$ and using it in eq. (\ref{sil2}) we
obtain
\begin{equation}
\frac{\left( 2n+1\right) \beta ^{2}}{4\pi d}\frac{c^{2}t^{-2-b}}{Gk_{\gamma
}^{-b}}\left[ \frac{\dot{c}}{c}+\frac{1}{t}\right] +\frac{b}{t}=0,
\label{cova}
\end{equation}
where $\beta =-b/\alpha .$ This is a Bernouilli ode and its
solution is:
\begin{equation}
c(t)^{2}=-\frac{1}{t^{2}}\frac{2bk_{\gamma }^{-b}}{\left( 2n+1\right) A}\int
G(t)t^{3+b}dt
\end{equation}
with $A=\frac{\beta ^{2}}{4\pi d}.$ We can see that the above equation
verifies the relationship
\begin{equation}
\frac{k_{\gamma }^{b}c^{2}}{G}=gt^{b+2}\text{ \ \ \ \ \ therefore \ }%
G=gk_{\gamma }^{b}c^{2}t^{-b-2},
\end{equation}
$g$ being a numerical constant. It is observed that if $b=-2$ i.e. $\gamma
=1/2$, we obtain $G/c^{2}=B$. We use this relation in eq.(\ref{cova})
obtaining
\begin{equation}
\frac{\dot{c}}{c}=\left( -1-\frac{b}{K_{c}}\right) t^{-1},
\end{equation}
where $K_{c}=\frac{\left( 2n+1\right) \beta ^{2}}{4\pi dg}$, and therefore,
\begin{equation}
c=\tilde{K}_{c}t^{\kappa },
\end{equation}
where $\tilde{K}_{c}$ is an integration constant and $\kappa =\left( -1-%
\frac{b}{K_{c}}\right) .$

Therefore the expansion and the shear yield
\begin{equation}
\theta :=\frac{1}{c(t)}\left( \frac{\dot{X}}{X}+2\frac{\dot{Y}}{Y}\right) =-%
\frac{b\left( n+2\right) }{\tilde{K}_{c}\alpha t^{\kappa +1}}
\end{equation}
and
\begin{equation}
\sigma =\frac{\sqrt{3}}{3c(t)}\left( \frac{\dot{X}}{X}-\frac{\dot{Y}}{Y}%
\right) =-\frac{\sqrt{3}}{3}\frac{b\left( n-1\right) }{\tilde{K}_{c}\alpha
t^{\kappa +1}}
\end{equation}
while the Kretschmann scalars are:
\begin{equation}
K_{1}\thickapprox C_{1}t^{-4\left( \kappa +1\right) },\text{ \ \ \ \ \ \ \ \
\ }K_{2}\thickapprox C_{2}t^{-4\left( \kappa +1\right) }
\end{equation}
we have omitted all the tedious calculations.

If we impose a new assumption (which is supported by observational
considerations of $\Lambda$ decaying with time and assuming a small but
non-zero value) as
\begin{equation}
\Lambda=\frac{l}{c^{2}(t)t^{2}}\text{ }\Longrightarrow\text{ \ }\dot{\Lambda
}=-\frac{2l}{c^{2}t^{2}}\left( \frac{\dot{c}}{c}+\frac{1}{t}\right) ,
\end{equation}
we can see from eq. (\ref{sil1}) that
\begin{equation}
\frac{G}{c^{2}}=\left[ \frac{3\beta^{2}-l}{8\pi d}\right] k_{%
\gamma}^{b}t^{-b-2}=Bk_{\gamma}^{b}t^{-b-2}.
\end{equation}
Taking all these relations into consideration, we use in eq.(\ref{sil2}) to
obtain
\begin{equation}
-\frac{l}{4\pi B}\left( \frac{\dot{c}}{c}+\frac{1}{t}\right) -2\frac{\dot {c}%
}{c}-\frac{b+2}{t}=0,
\end{equation}
whose trivial solution is
\begin{equation}
c=\mathfrak{K}_{c}t^{-\mu},
\end{equation}
where $\mathfrak{K}_{c}$ is an integration constant and $\mu=\left( \frac{%
l+4\pi K(b+2)}{l+8\pi K}\right) .$

\subsubsection{Conclusions for this model.}

We have solved this model under the assumptions: $div(T)=0,$ $\Pi =\varkappa
\rho $, with $\varkappa \in \mathbb{R}^{-}$ and valid for $\forall \gamma .$
It is observed that if $\gamma =1/2$ we obtain $G/c^{2}=const.$ as obtained
as integration condition in our previous paper \cite{Tony1}.

In the considered model the evolution of the Universe starts from a singular
state, as the Krestchmann scalars show us, with the energy density, bulk viscosity coefficient and cosmological
constant tending to infinity. At the initial moment $t=0$ the relaxation time
is $\tau=0$. Generally the behavior of the gravitational constant shows a
strong dependence on the coefficient $\gamma$ entering in the equation of state of
the bulk viscosity coefficient. For $\gamma=1/2$, $G/c^{2}$ is a constant during the
cosmological evolution. From the singular initial state the Universe starts to
expand, with the scale factors $X$ and $Y$ as a $\gamma$-dependent functions. Depending on the
value of the constants in the equations of state, the cosmological evolution
is non-inflationary ($q>0$ for $H_{0}<-1$) or inflationary ($q<0$ for $H_{0}%
>1$). Due to the proportionality between bulk viscous pressure and energy
density, both inflationary and non-inflationary models are thermodynamically
consistent, with the ratio of $\Pi$ and $p$ also much smaller than $1$ in the
inflationary era. Generally, the cosmological constant is a decreasing
function of time.

We also notice that the $\varkappa -$parameter, i.e. the causal bulk viscous
effect, weakly perturbs the FRW perfect fluid solution in such a way that
the comoving entropy varies with $t,$ while in the perfect fluid case, i.e. $%
\varkappa =0$, the comoving entropy is constant. The same considerations
could be taken into account for the scale factor $Y(t)=Y_{0}t^{\frac{1}{%
3\left( 1-\gamma \right) \left( \omega +1+\varkappa \right) \left(
n+2\right) }},$ which is perturbed weakly for the bulk viscous parameter as
we can see in the special case of $\gamma =1/2$, and for the $n-$parameter$.$
The solutions obtained for the scale factor, energy density, entropy, etc.
are similar ``but different'' to those obtained for a perfect fluid model
where the $\varkappa -$parameter vanishes. We would like to point out that
our model is thermodynamically consistent for the usual matter equations of
state and valid for all $\gamma -$parameter, i.e. $\forall \gamma \in \left(
0,1\right) .$ In the same way, we can see that the viscous parameter helps
us to get rid of the so-called entropy problem since in this model entropy
varies with $t.$

\subsection{Model 2. Hypotheses 2a.}

If we follow the hypothesis $2a$ then it is observed that
\begin{equation}
Y=\left( C_{2}(n+2)+C_{1}(n+2)\int c(t)dt\right) ^{\frac{1}{(n+2)}}
\end{equation}
and making
\begin{equation}
c(t)=c_{0}t^{a}
\end{equation}
with $a\in \mathbb{R},$ therefore
\begin{equation}
Y=\left( C_{2}(n+2)+C_{1}\frac{(n+2)}{a+1}t^{a+1}\right) ^{\frac{1}{(n+2)}},%
\text{ \ \ \ \ }\Longrightarrow \text{\ \ \ }X=\left( C_{2}(n+2)+C_{1}\frac{%
(n+2)}{a+1}t^{a+1}\right) ^{\frac{n}{(n+2)}}
\end{equation}
defining
\begin{equation}
H=\left( \frac{\dot{X}}{X}+2\frac{\dot{Y}}{Y}\right) =\frac{\left(
a+1\right) C_{1}t^{a}}{\left( C_{2}\left( a+1\right)
+C_{1}t^{a+1}\right) }
\end{equation}
we can calculate the deceleration parameter $q$%
\begin{equation}
q=\frac{d}{dt}\left( \frac{3}{H}\right) -1=3\left( -a\frac{C_{2}}{C_{1}}%
t^{-a-1}+\frac{1}{a+1}\right) -1
\end{equation}
as we can see this expression only has sense if $C_{2}=0,$ therefore
\begin{equation}
q=\frac{2-a}{a+1}  \label{q2a}
\end{equation}

Now taking into account the expression
\begin{equation}
\dot{\rho}+\left( \rho +p+\Pi \right) H=0,
\end{equation}
we take $\Pi $ as
\begin{equation}
\Pi =-\alpha \rho -\frac{\dot{\rho}}{H}
\end{equation}
$\alpha =\left( \omega +1\right) ,$ now simplify this expression into
\begin{equation}
\dot{\Pi}+\frac{\Pi }{k_{\gamma }\rho ^{\gamma -1}}+\rho H=-\frac{1}{2}\Pi
\left( H+W\frac{\dot{\rho}}{\rho }\right)
\end{equation}
with $W=\left( \frac{2\omega +1}{\omega +1}\right) ,$ we obtain a second
order ode for $\rho (t).$%
\begin{equation}
\ddot{\rho}=\left( \frac{\dot{H}}{H}-\frac{\rho ^{1-\gamma }}{k_{\gamma }}%
-AH\right) \dot{\rho}-BH^{2}\rho +CH\rho ^{2-\gamma }+D\frac{\dot{\rho}^{2}}{%
\rho }
\end{equation}
where $A=\left( \frac{1}{2}+\frac{\alpha W}{2}+\alpha \right) ,$ $B=\left( 1-%
\frac{\alpha }{2}\right) ,$ $C=\left( \frac{\alpha }{k_{\gamma }}\right) ,$ $%
D=\frac{W}{2},$ simplifying it is obtained:
\begin{equation}
\ddot{\rho}=\left( \frac{a}{t}-\left( A+1\right) H-\frac{\rho ^{s}}{%
k_{\gamma }}\right) \dot{\rho}-BH^{2}\rho +CH\rho ^{s+1}+D\frac{\dot{\rho}%
^{2}}{\rho }  \label{paula}
\end{equation}
(where $s=1-\gamma )$ which is a second order differential equation with
linear symmetries.

Taking into account the standard Lie procedure \cite{59, 62} to obtaining the symmetries
of this ode we see that (\ref{paula}) only admits the following symmetry
\begin{equation}
\xi =-st,\text{ \ \ \ \ }\eta =\rho
\end{equation}
iff $C_{2}=0,$ (as we already know) obtaining in this way the following
invariant solution:
\begin{equation}
\frac{dt}{\left( \gamma -1\right) \left( t\right) }=\frac{d\rho }{\rho }%
\Longrightarrow \rho =d_{0}t^{b},\text{ \ \ \ \ \ \ \ \ \ }b=\frac{1}{\gamma
-1}
\end{equation}
therefore
\begin{equation}
\Pi =-\alpha \rho -\frac{\dot{\rho}}{H}\Longrightarrow \Pi =\Pi _{0}t^{b},%
\text{ \ \ \ \ \ \ \ \ }\Pi _{0}=d_{0}k_{\gamma }^{-b}\left( -\alpha -\frac{b%
}{a+1}\right)
\end{equation}
it is observed that we have obtained the same behaviour (in order of
magnitude) than the energy density, showing in this way that our first
hypothesis at least has mathematical meaning. We observe that this solution
only has physical sense if $a>0,$ since $\Pi <0.$

The entropy is
\begin{equation}
\Sigma (t)-\Sigma \left( t_{0}\right) =\Sigma _{0}\left[ \tilde{t}%
^{b(1-\delta )+(a+1)}\right] _{t_{0}}^{t}  \label{E2a}
\end{equation}
if $a>0,$ then $\Sigma $ is a growing function on time $t.$ With these
restriction we find that $q$ behaves as:
\begin{equation}
q=\frac{2-a}{a+1}<0\text{ \ \ }\Longleftrightarrow \text{\ \ \ \ }a>2,
\end{equation}
finding that this universe accelerates.

Now from eq.
\begin{equation}
\left( 2n+1\right) \frac{\dot{Y}^{2}}{Y^{2}}=\frac{8\pi G}{c^{2}}\rho
+\Lambda c^{2},
\end{equation}
and defining
\begin{equation}
\left( 2n+1\right) \frac{\dot{Y}^{2}}{Y^{2}}=f(t)
\end{equation}
we obtain $\Lambda $%
\begin{equation}
\left( f(t)-\frac{8\pi G}{c^{2}}\rho (t)\right) \frac{1}{c^{2}(t)}=\Lambda
\label{lam1}
\end{equation}
and substituting this expression into
\begin{equation}
\frac{\dot{\Lambda}c^{4}}{8\pi G\rho }+\frac{\dot{G}}{G}-4\frac{\dot{c}}{c}%
=0.
\end{equation}
we end obtaining the behaviour of the ``constant'' $G$ as:
\begin{equation}
G=\frac{c^{2}f}{8\pi \dot{\rho}}\left( \frac{\dot{f}}{f}-2\frac{\dot{c}}{c}%
\right)
\end{equation}
therefore it yields
\begin{equation}
G=G_{0}t^{2(a-1)-b}  \label{G2a}
\end{equation}
it is observed that we have again the relationship
\begin{equation}
\frac{G}{c^{2}}\thickapprox t^{-2-b},\text{ \ \ \ \ \ \ \ \ }\frac{G}{c^{2}}%
=const.\Longleftrightarrow b=-2\Longleftrightarrow \gamma =1/2.
\end{equation}

Once it has been obtained $G$ then we back to eq. (\ref{lam1}) obtaining in
this way the behaviour of $\Lambda :$%
\begin{equation}
\Lambda =\Lambda _{0}t^{-2(a+1)}  \label{L2a}
\end{equation}

The behaviour of
\begin{equation}
K_{1}\thickapprox t^{-4(a+1)},\text{ \ \ \ \ \ \ \ \ \ \ \ }%
K_{2}\thickapprox t^{-4(a+1)},
\end{equation}
while the expansion and the shear behaves as:
\begin{equation}
\theta =\frac{a+1}{K}t^{-(a+1)},
\end{equation}
\begin{equation}
\sigma =\frac{\sqrt{3}}{3}\frac{\left( n-1\right) }{\left( n+2\right) }%
\left( a+1\right) t^{-(a+1)}.
\end{equation}

\subsubsection{Conclusion for this model. Numerical values and graphics for
the main quantities.}

In the first place we begin studying the behaviour of the radius of our
model. For this purpose we need to fix some numerical constants as well as
consider different equations of state.

In the rest of this section we shall consider the following values for the
numerical constants
\begin{equation*}
C_{1}=1,\quad C_{2}=0,\quad c_{0}=1,\quad d_{0}=1,\quad
b=-2\Longleftrightarrow \gamma =\frac{1}{2}
\end{equation*}
and
\begin{equation*}
\begin{array}{c|c|c|c}
a & n & \omega & color \\ \hline
\frac{1}{2} & \frac{3}{2} & 1 & red \\ \hline
\frac{1}{3} & \frac{1}{2} & \frac{1}{3} & blue \\ \hline
1 & \frac{1}{3} & 0 & magenta \\ \hline
\frac{-1}{2} & \frac{1}{2} & 1 & black
\end{array}
\end{equation*}
we have chosen the last values (black color) as pathological.

\begin{figure}[h]
\begin{center}
\includegraphics[height=2.194in,width=2.194in]{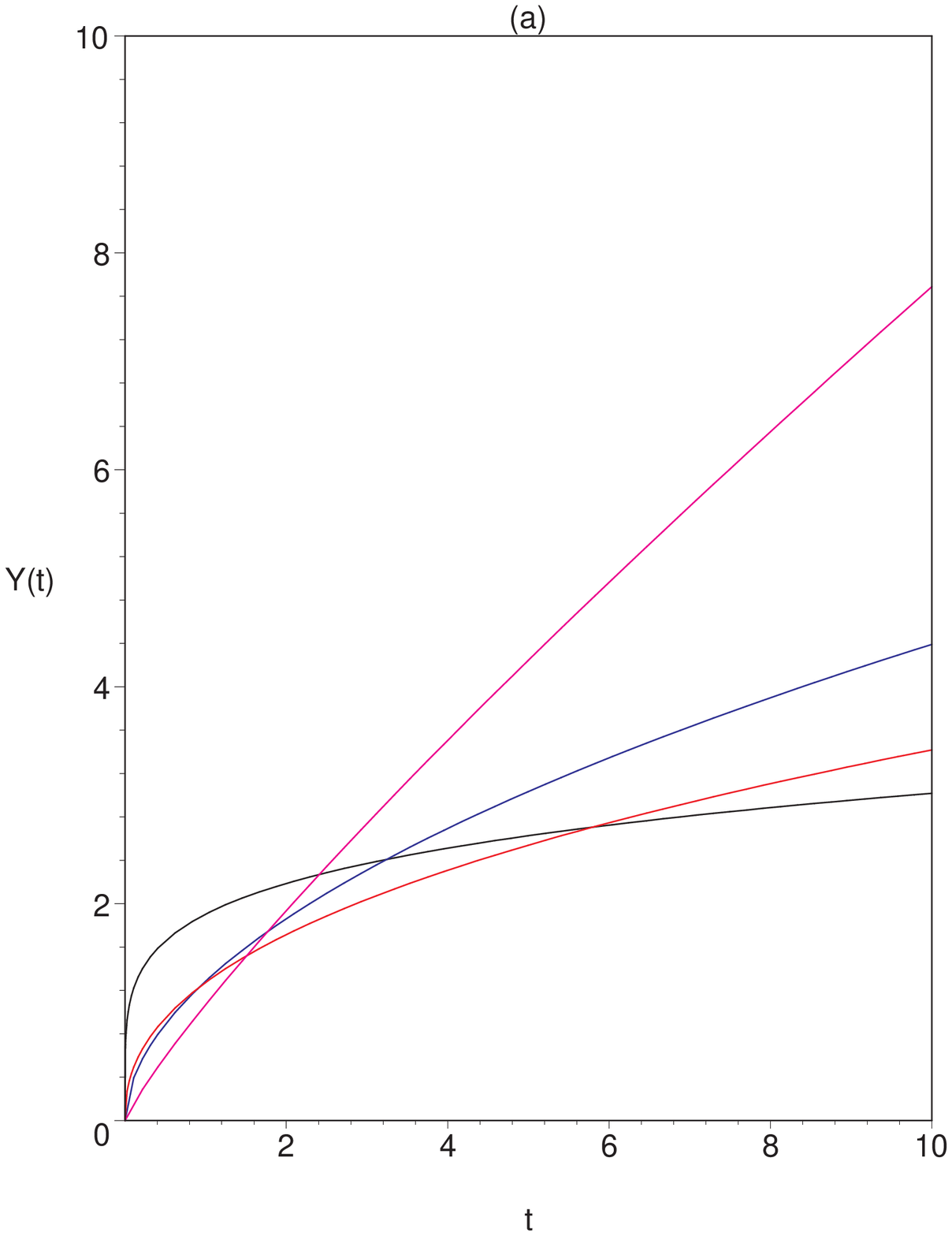} %
\includegraphics[height=2.194in,width=2.194in]{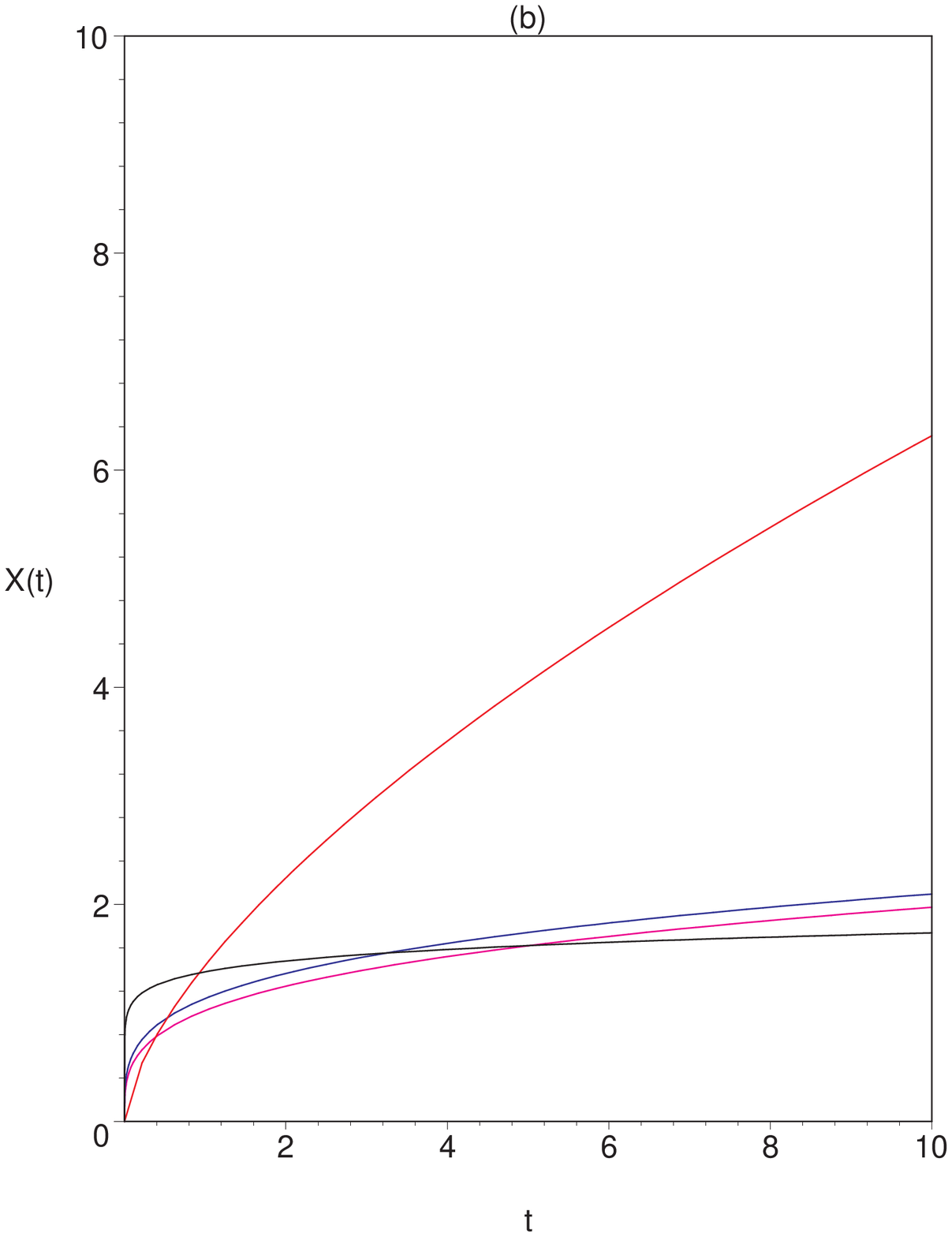}
\end{center}
\caption{In the figure (a) it is plotted the radius $Y(t)$ while in figure (b) it is plotted the variation of the radius $X(t)$. Both radius have a singular start and are growing functions on time.}
\label{Radios2}
\end{figure}

With these numerical values we can see in fig. (\ref{Radios2}) the different
behaviour of our scale factors. In all the studied cases we can see that
these scale factors are growing functions on time $t$. These solutions are
singular as the Krestchmann invariants show us.

\begin{figure}[h!]
\begin{center}
\includegraphics[height=2.194in,width=2.194in]{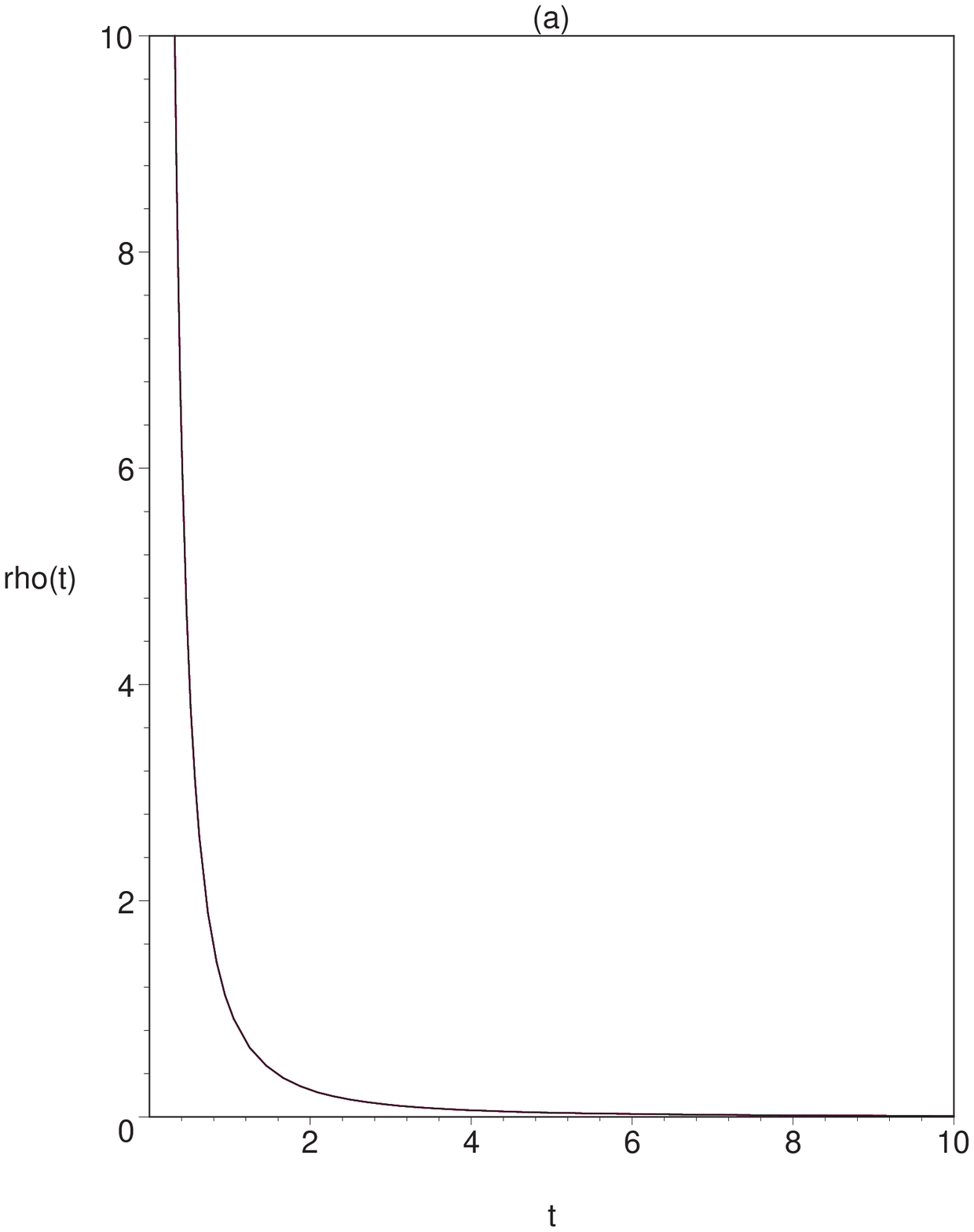} %
\includegraphics[height=2.194in,width=2.194in]{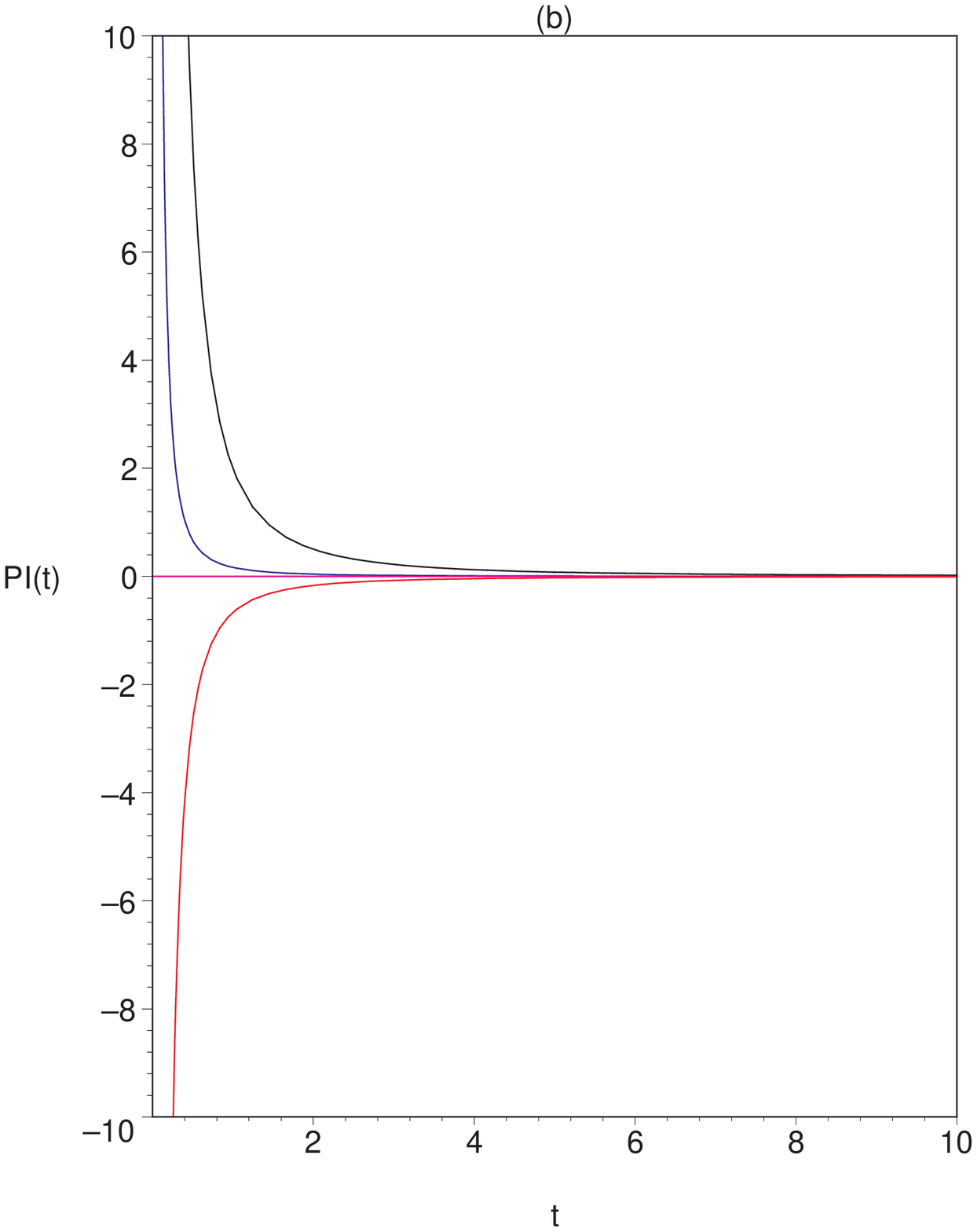}
\end{center}
\caption{Figure (a) shows the variation of the energy density $\protect%
\rho(t)$. Figure (b) shows the variation of the bulk viscous pressure $%
\Pi(t)$.}
\label{densidad2}
\end{figure}

With regard to the energy density and the bulk viscous pressure
(see fig. (\ref {densidad2})) we can see that all the solutions
have no physical meaning except the case
$a=\frac{1}{2},n=\frac{3}{2}$ and $\omega =1$ (ultrastiff matter)
which correspond to the red plot, since this is the only solution
that verifies the condition $\Pi<0$ for all time interval and for
all values of the parameters. It is observed that for $\omega =0$
(magenta line) $\Pi =0,$ i.e. $\Pi$ vanishes. Since we have only
been able to obtain a particular solution (invariant solution) to
the differential equation (\ref{paula}) for the energy density
then this quantity shows the same behaviour for the chosen
numerical constants.

\begin{figure}[h!]
\begin{center}
\includegraphics[height=2.194in,width=2.194in]{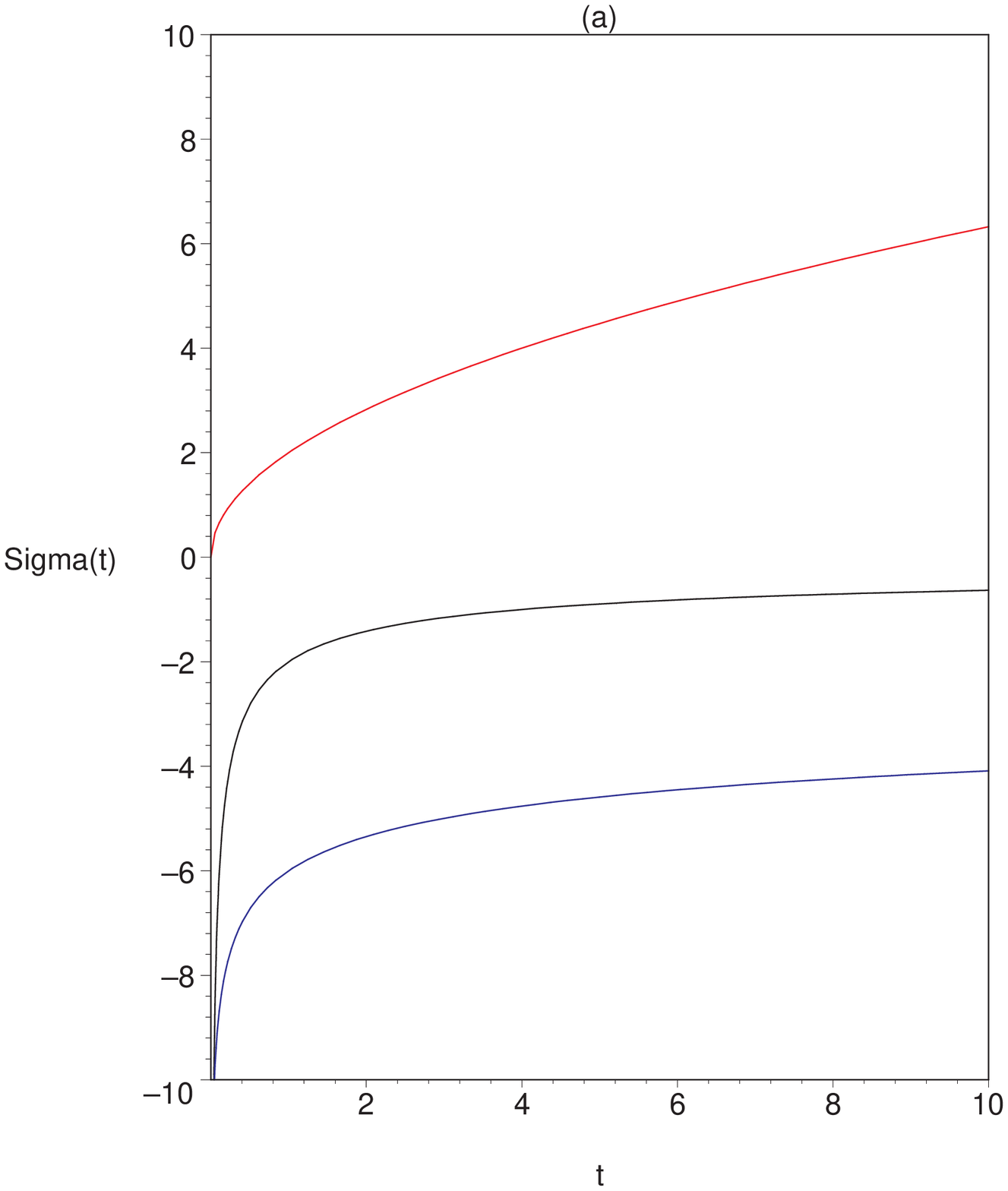} %
\includegraphics[height=2.194in,width=2.194in]{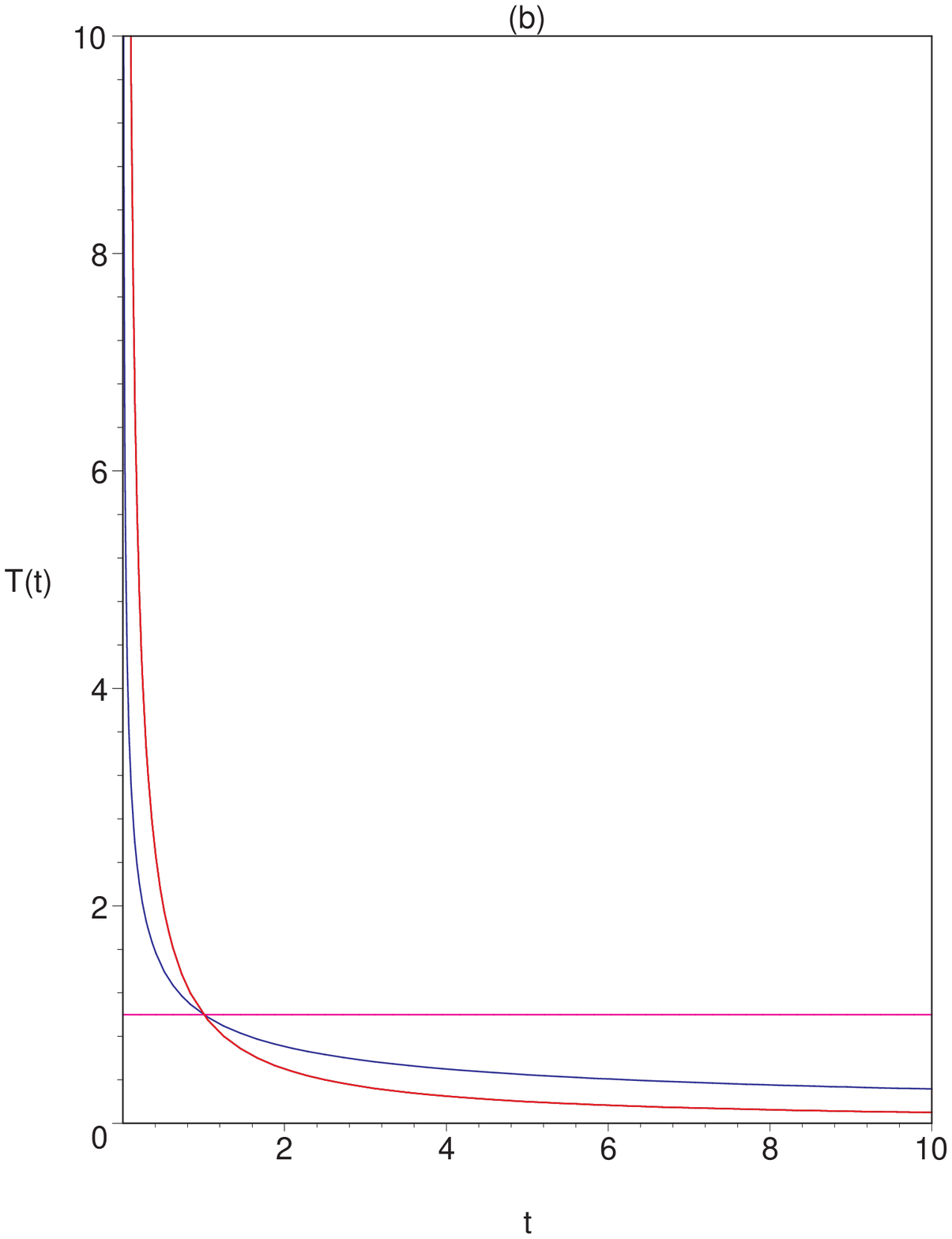}
\end{center}
\caption{Figure (a)  shows the variation of the entropy $\Sigma (t)$ while figure
 (b)  shows the variation of the temperature $T(t)$.}
\label{termo2}
\end{figure}

The behaviour of the thermodynamical quantities is showed in fig.
(\ref {termo2}). Even though all the studied cases show a
decreasing temperature except the case $\omega =0$, magenta line,
which correspond to $T(t)=const.,$ as it is expected, figure (a)
shows us that only the red solution has physical sense as already
we know from the last picture of the viscous pressure. This case
corresponds to a growing entropy.

\begin{figure}[h!]
\begin{center}
\includegraphics[height=2.194in,width=2.194in]{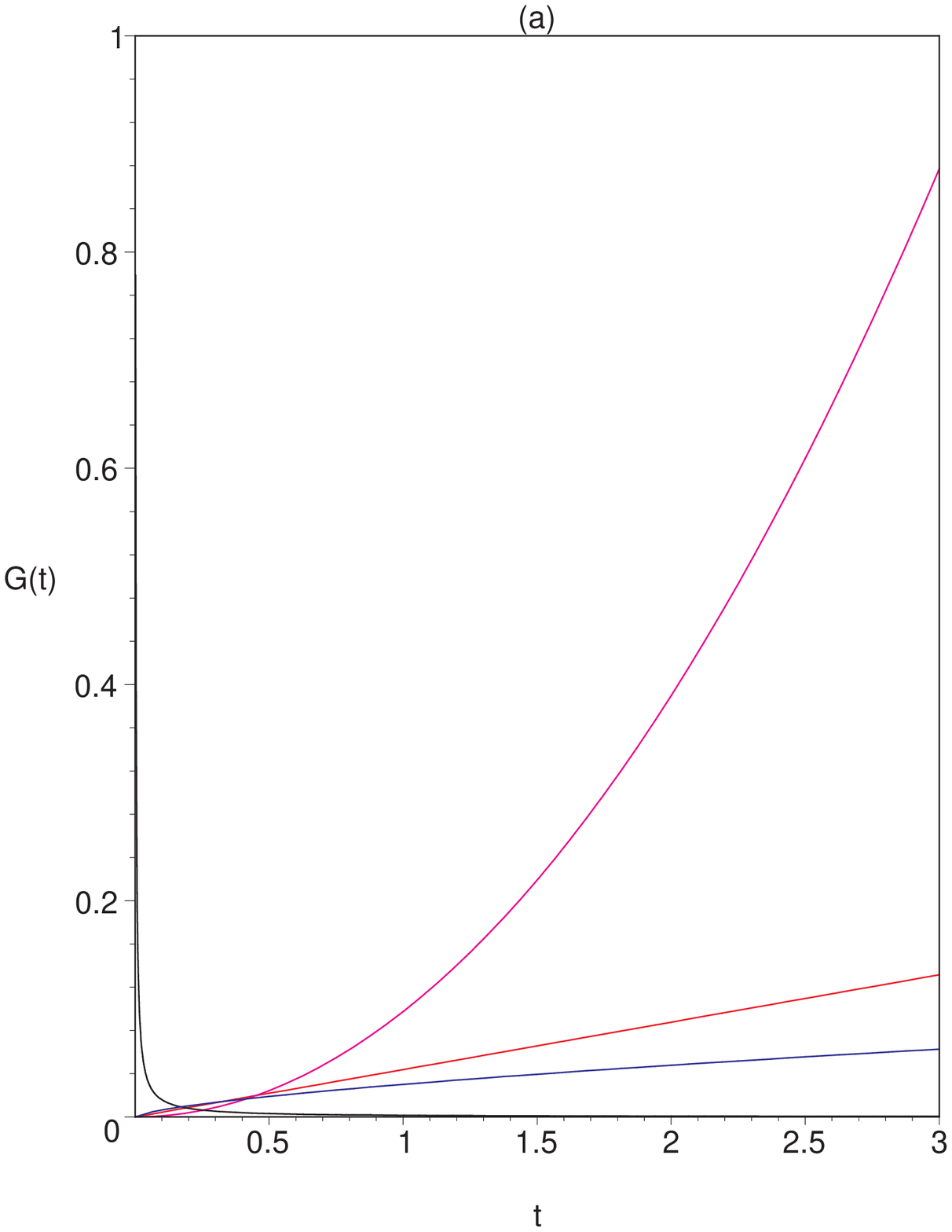} %
\includegraphics[height=2.194in,width=2.194in]{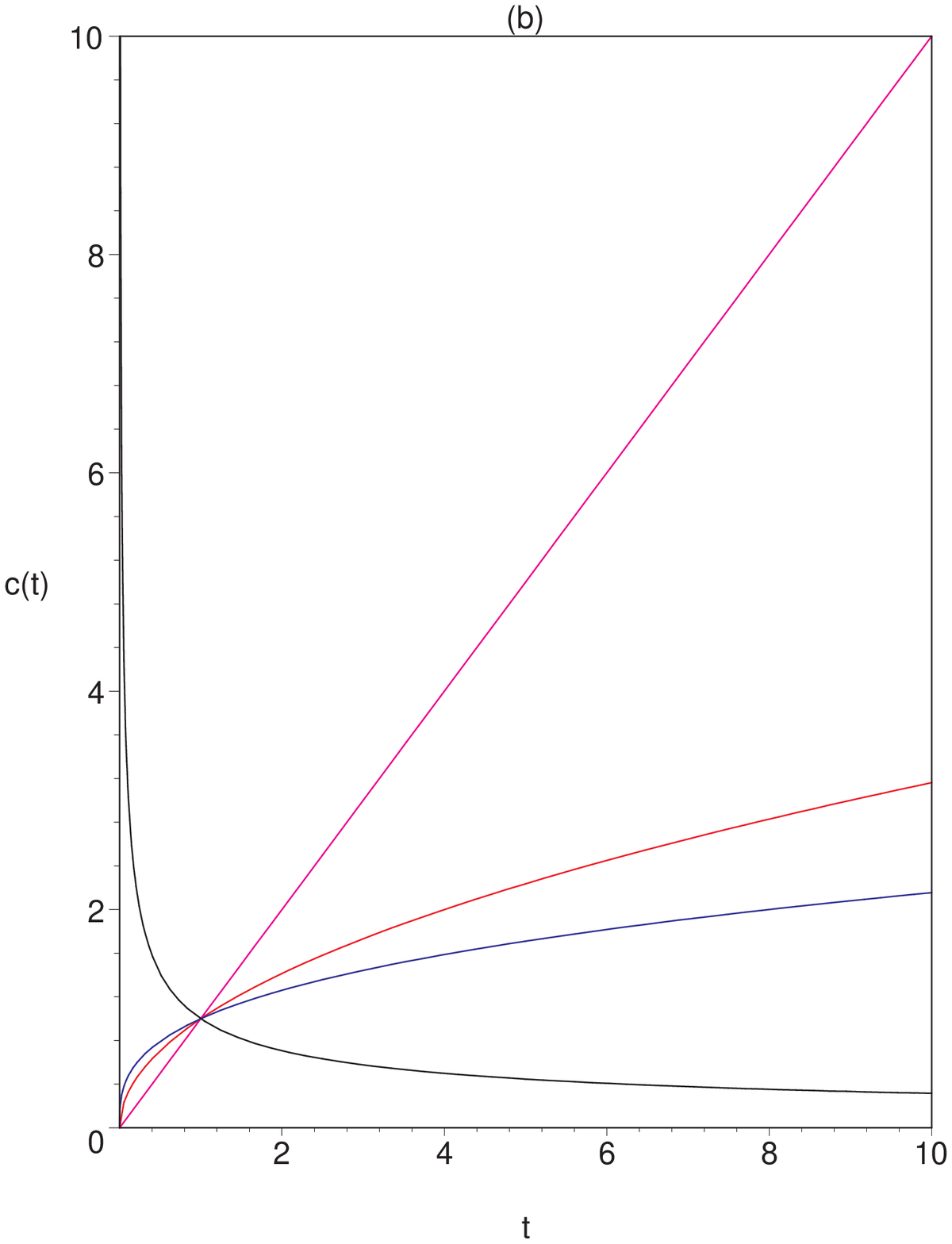} %
\includegraphics[height=2.194in,width=2.194in]{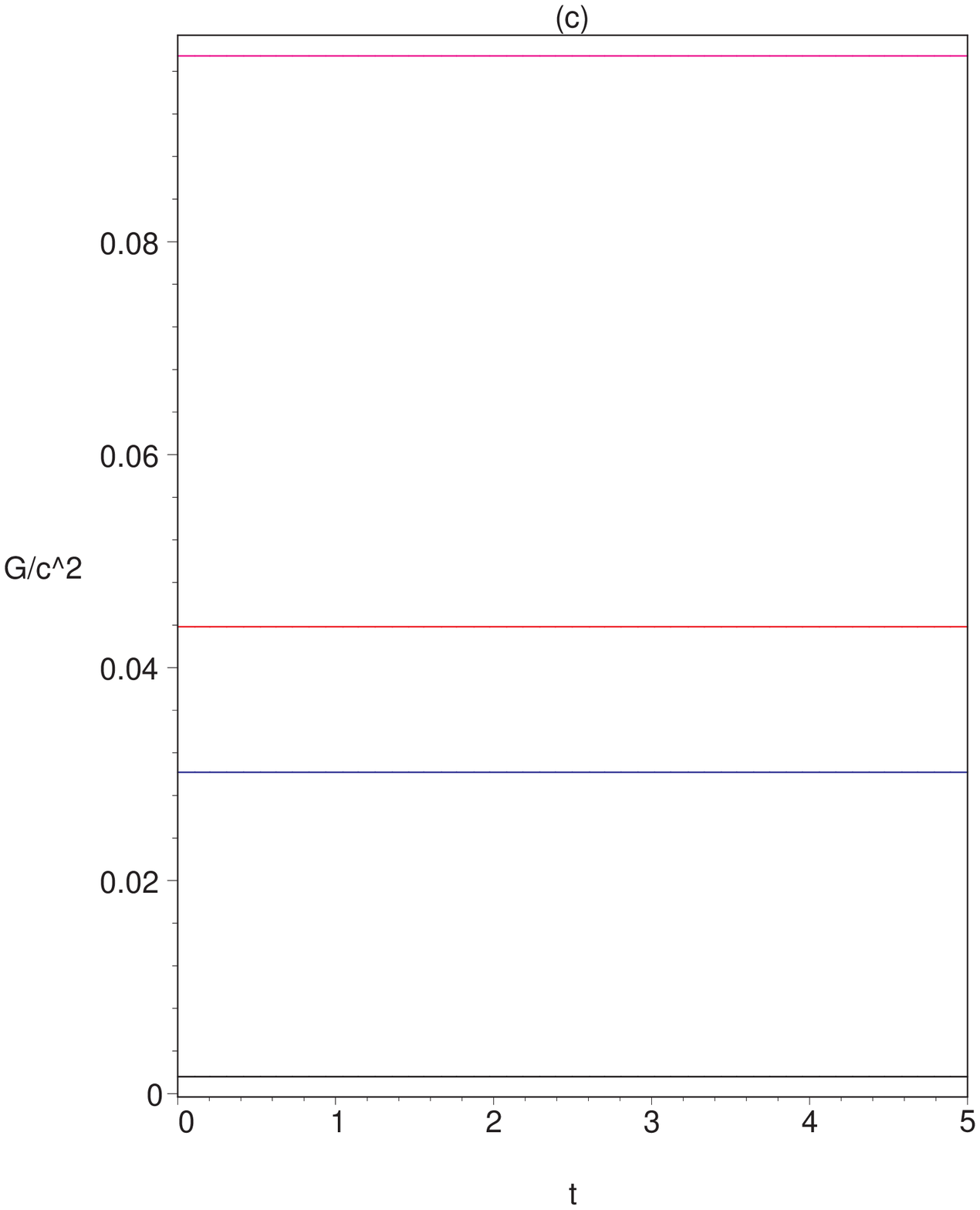}
\end{center}
\caption{In the figure (a) we show the variation of ``constant" $G(t)$. In the figure
 (b) we show the variation of the ``constant" $c(t)$ and in the figure (c) it is plotted the relationship
 $G/c^{2}$.}
\label{constantes2}
\end{figure}

The variation of the ``constants'' $G$ and $c$ as well as the relationship $%
G/c^{2}$ is shown in fig. (\ref{constantes2}). This figures show us that
both ``constants'' are growing functions on time $t$ except in the case
(black line) $n=-1/2$ where $G$ and $c$ are a decreasing functions on time $t$.
We would like to emphasize that only for the case $%
\gamma =1/2$ we have that it is verified the relationship
$G/c^{2}=const.$ It is a very plausible hypothesis that these
effects were much stronger in the early Universe, when dissipative
effects also played an important role in the dynamics of the
cosmological fluid. Hence the solution obtained (red line) could
give an appropriate description of the early period of our
Universe.

\begin{figure}[h!]
\begin{center}
\includegraphics[height=2.194in,width=2.194in]{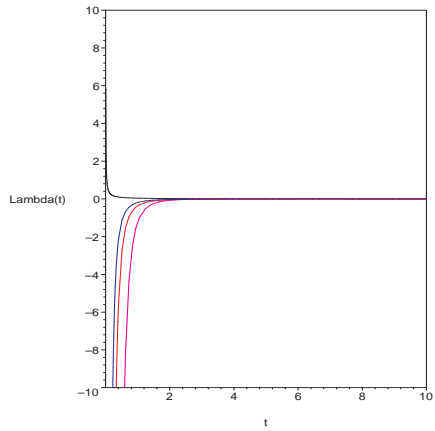}
\end{center}
\caption{The variation of the cosmological ``constant"  $\Lambda(t)$.}
\label{lambda2}
\end{figure}

With regard to the cosmological ``constant'', see fig. (\ref{lambda2}), it
is observed that all the solutions are decreasing but negative except for
the black line ($n=-1/2).$ Maybe this fact tells us that we may consider $\Lambda(t)$ as a true ghost energy density.

\begin{figure}[h]
\begin{center}
\includegraphics[height=2.194in,width=2.194in]{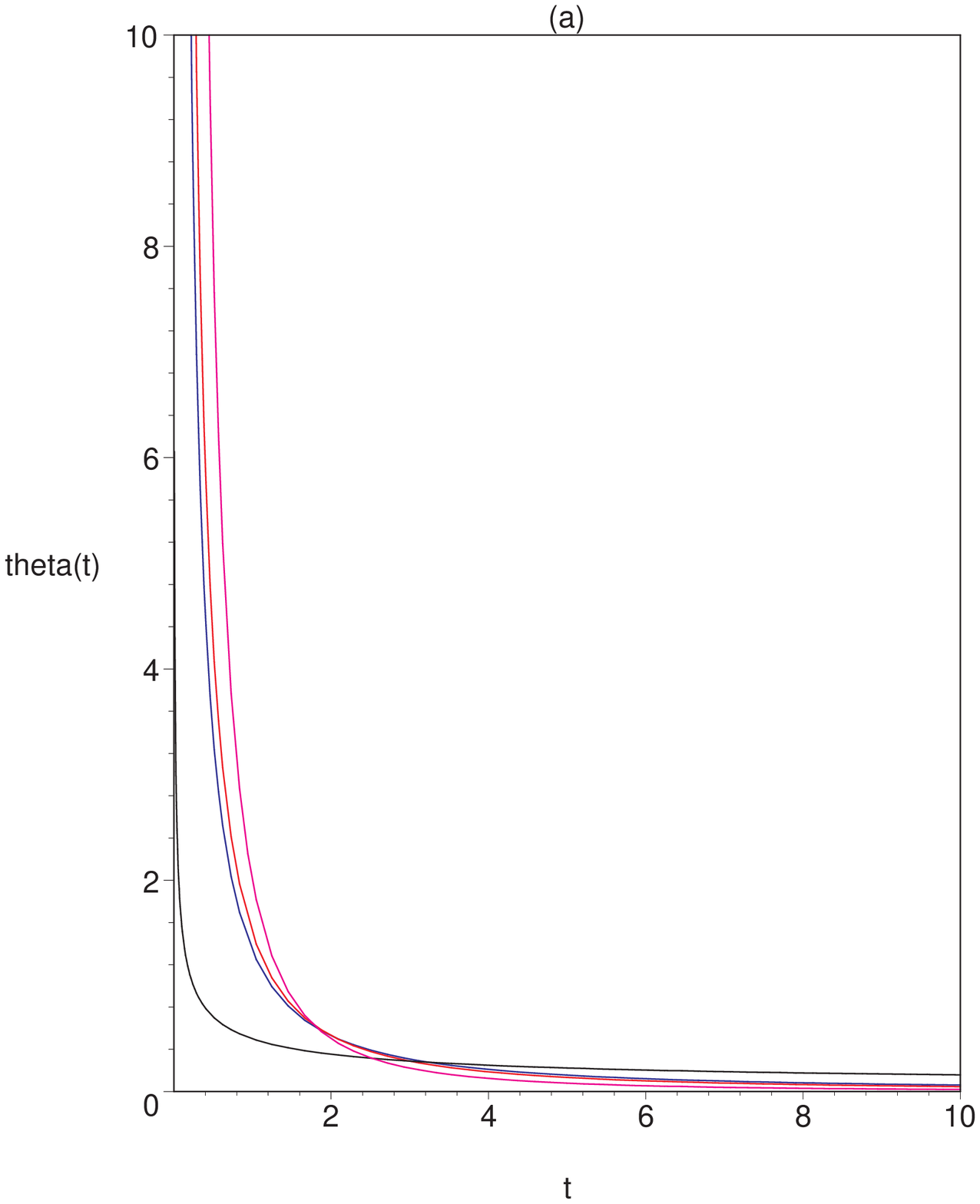} %
\includegraphics[height=2.194in,width=2.194in]{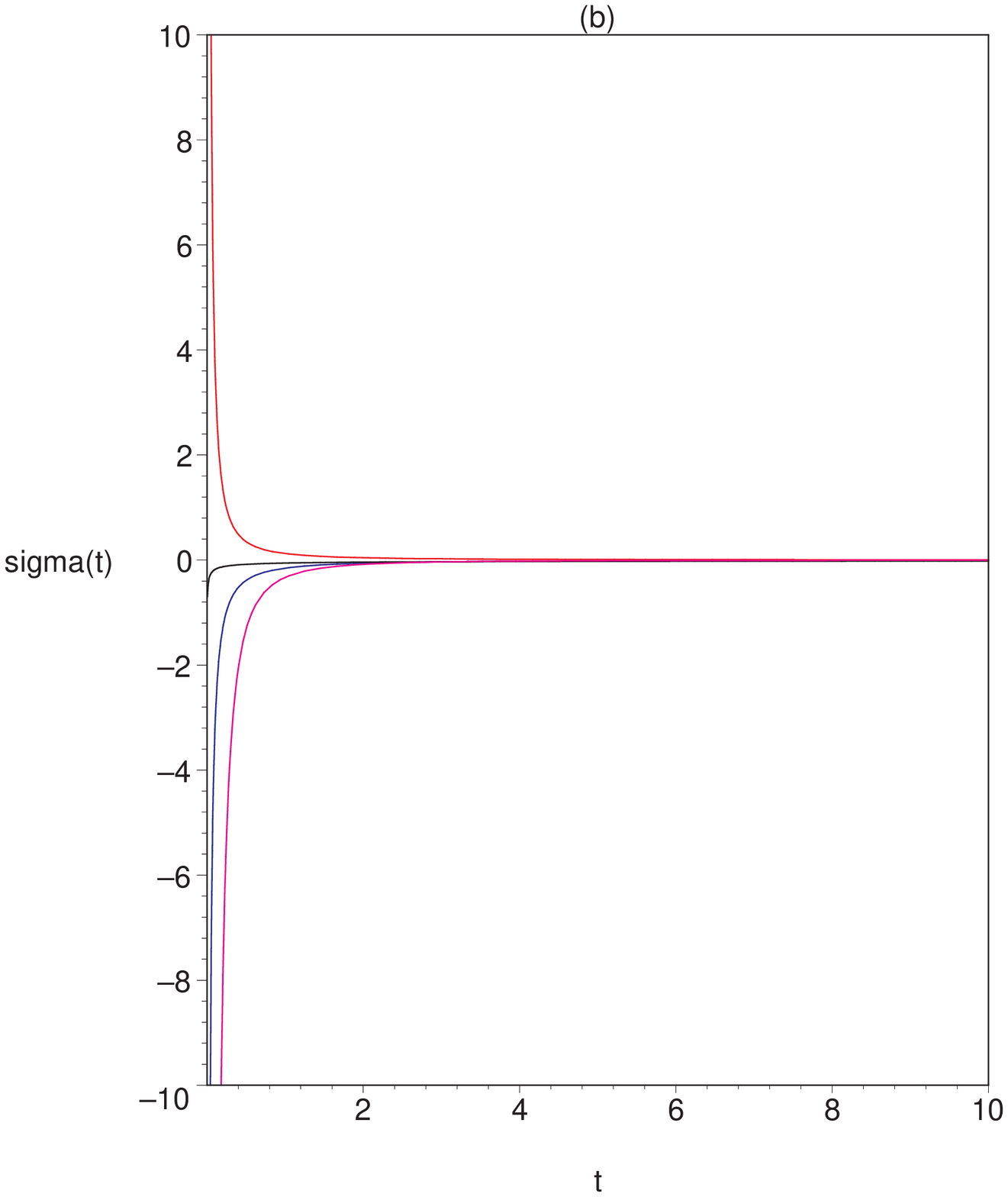}
\end{center}
\caption{In figure (a) it is plotted the expansion $\protect\theta (t)$,
while in figure (b) it is plotted the shear $\protect\sigma (t)$.}
\label{expansion2}
\end{figure}

The expansion and the shear behave as follows, see fig.
(\ref{expansion2}). As we can see all the models studied show a
decreasing expansion and the only model that has a positive shear
is the plotted with the red color. The shear and expansion scalars
calculated above indicate that the Universe is shearing and
expanding with time. For the red color solution, the shear, which
is a degree of anisotropy in the Universe, decreases monotonically
with time. This indicates the fact that the initially anisotropic
Universe gradually tends to an isotropic Universe at late time
(present epoch) which is in agreement with the recent observations
of a negligible amount of anisotropy present in the CMBR.

Other solutions can be studied changing the numerical values of the
constants as well as taking into account different equations of state. We
have found at least a physical solution (red line) which equations of state
are $\omega =1$ (ultrastiff \ matter) and $\gamma =1/2$ (the standard value
for the viscous parameter).

\subsection{Model 3. Hypotheses 2b.}

If we follow the hypothesis $2b$ then it is observed that if we make the
next assumption
\begin{equation}
c(t)=c_{0}Y^{n}
\end{equation}
then eq. (\ref{M_T}) has the following solution:
\begin{equation}
Y=X\left( C_{2}\left( n-2\right) \int X^{n-3}dt+C_{1}(2-n)\right) ^{\frac{1}{%
2-n}}
\end{equation}
then we will need to make a hypothesis about the behaviour of $X(t)$ in
order to obtain a complete solution. Now if we assume that
\begin{equation}
X=Y^{m}
\end{equation}
then
\begin{equation}
Y=\left( C_{1}\left( m-n+2\right) t+C_{2}\left( m-n+2\right) \right) ^{\frac{%
1}{\left( m-n+2\right) }}
\end{equation}
defining
\begin{equation}
H=\left( \frac{\dot{X}}{X}+2\frac{\dot{Y}}{Y}\right) =\frac{\left(
m+2\right) C_{1}}{\left( m-n+2\right) \left( C_{1}t+C_{2}\right) }
\end{equation}
we can calculate the deceleration parameter $q$%
\begin{equation}
q=\frac{d}{dt}\left( \frac{3}{H}\right) -1=3-\frac{3n}{m+2}.
\end{equation}

Now taking into account the expression
\begin{equation}
\dot{\rho}+\left( \rho +p+\Pi \right) H=0,
\end{equation}
we take $\Pi $ as
\begin{equation}
\Pi =-\alpha \rho -\frac{\dot{\rho}}{H}
\end{equation}
$\alpha =\left( \omega +1\right) ,$ now simplify this expression into
\begin{equation}
\dot{\Pi}+\frac{\Pi }{k_{\gamma }\rho ^{\gamma -1}}+\rho H=-\frac{1}{2}\Pi
\left( H+W\frac{\dot{\rho}}{\rho }\right)
\end{equation}
with $W=\left( \frac{2\omega +1}{\omega +1}\right) ,$ we obtain a second
order ode for $\rho (t).$%
\begin{equation}
\ddot{\rho}=\left( \frac{\dot{H}}{H}-\frac{\rho ^{1-\gamma }}{k_{\gamma }}%
-AH\right) \dot{\rho}-BH^{2}\rho +CH\rho ^{2-\gamma }+D\frac{\dot{\rho}^{2}}{%
\rho }
\end{equation}
where $A=\left( \frac{1}{2}+\frac{\alpha W}{2}+\alpha \right) ,$ $B=\left( 1-%
\frac{\alpha }{2}\right) ,$ $C=\left( \frac{\alpha }{k_{\gamma }}\right) ,$ $%
D=\frac{W}{2},$ simplifying it is obtained:
\begin{equation}
\ddot{\rho}=-\left( \frac{C_{1}}{C_{1}t+C_{2}}+AH+\frac{\rho ^{s}}{k_{\gamma
}}\right) \dot{\rho}-BH^{2}\rho +CH\rho ^{s+1}+D\frac{\dot{\rho}^{2}}{\rho }
\label{claudia}
\end{equation}
(where $s=1-\gamma )$ which is a second order differential equation with
linear symmetries.

Taking into account the standard Lie procedure \cite{59, 62} to obtaining the symmetries
of this ode we see that (\ref{claudia}) only admits the following symmetry
\begin{equation}
\xi =-s\left( C_{1}t+C_{2}\right) ,\text{ \ \ \ \ }\eta =\rho
\end{equation}
obtaining in this way the following invariant solution:
\begin{equation}
\frac{dt}{\left( \gamma -1\right) \left( C_{1}t+C_{2}\right) }=\frac{d\rho }{%
\rho }\Longrightarrow \rho =d_{0}\left( C_{1}t+C_{2}\right) ^{b}
\end{equation}
therefore
\begin{equation}
\Pi =-\alpha \rho -\frac{\dot{\rho}}{H}\Longrightarrow \Pi =\Pi _{0}\left(
C_{1}t+C_{2}\right) ^{b}
\end{equation}
where $\Pi _{0}=-d_{0}\left( \alpha +b\left( 1-\frac{n}{m+2}\right) \right)
. $ It is observed \ the we have obtained the same behaviour (in order of
magnitude) than the energy density.

The entropy is
\begin{equation}
\Sigma (t)-\Sigma \left( t_{0}\right) =\Sigma _{0}\left[ \left( C_{1}\tilde{t%
}+C_{2}\right) ^{b(1-\delta )+(\frac{m+2}{m-n+2})}\right] _{t_{0}}^{t}
\label{E2b}
\end{equation}

Now from
\begin{equation}
\left( 2n+1\right) \frac{\dot{Y}^{2}}{Y^{2}}=\frac{8\pi G}{c^{2}}\rho
+\Lambda c^{2},
\end{equation}
where
\begin{equation}
\left( 2n+1\right) \frac{\dot{Y}^{2}}{Y^{2}}=f(t)
\end{equation}
we obtain $\Lambda $%
\begin{equation}
\left( f(t)-\frac{8\pi G}{c^{2}}\rho (t)\right) \frac{1}{c^{2}(t)}=\Lambda
\label{lam2}
\end{equation}
and substituting this expression into
\begin{equation}
\frac{\dot{\Lambda}c^{4}}{8\pi G\rho }+\frac{\dot{G}}{G}-4\frac{\dot{c}}{c}%
=0.
\end{equation}
we end obtaining the behaviour of the ``constants''.
\begin{equation}
\frac{c^{2}f}{8\pi G\rho }\left( \frac{\dot{f}}{f}-2\frac{\dot{c}}{c}\right)
-\frac{\dot{\rho}}{\rho }=0
\end{equation}
\begin{equation}
G=\frac{c^{2}f}{8\pi \dot{\rho}}\left( \frac{\dot{f}}{f}-2\frac{\dot{c}}{c}%
\right) =G_{0}\left( C_{1}t+C_{2}\right) ^{-2-b+\frac{2n}{m-n+2}}
\end{equation}
with $G_{0}>0.$ It is observed that we have again the relationship
\begin{equation}
\frac{G}{c^{2}}\thickapprox t^{-2-b},\text{ \ \ \ \ \ \ \ \ }\frac{G}{c^{2}}%
=const.\Longleftrightarrow b=-2\Longleftrightarrow \gamma =1/2.
\end{equation}

Once it has been obtained $G$ then we back to eq. (\ref{lam2}) obtaining in
this way the behaviour of $\Lambda :$%
\begin{equation}
\Lambda =\Lambda _{0}\left( C_{1}t+C_{2}\right) ^{\frac{-2(m+2)}{m-n+2}}
\end{equation}
with $\Lambda _{0}=const.$

The behaviour of the Krestchmann scalars are:
\begin{equation}
K_{1}\thickapprox \left( C_{1}t+C_{2}\right) ^{-\frac{4\left( n+1\right) }{%
m-n+2}},\text{ \ \ \ \ \ \ \ \ \ \ \ }K_{2}\thickapprox \left(
C_{1}t+C_{2}\right) ^{-\frac{4\left( n+1\right) }{m-n+2}},
\end{equation}
while the expansion and the shear behaves as:
\begin{equation}
\theta =C_{1}(m+2)\left( C_{1}t+C_{2}\right) ^{^{-\frac{m+2}{m-n+2}}},
\end{equation}
\begin{equation}
\sigma =\frac{\sqrt{3}}{3}C_{1}\left( m-1\right) \left( C_{1}t+C_{2}\right)
^{^{-\frac{m+2}{m-n+2}}}.
\end{equation}

\subsubsection{Conclusion for this model. Numerical values and graphics for
the main quantities.}

As in the above case we begin fixing the values of numerical constants and
the equation of state. In this occasion we take for the rest of this section
the following values for the numerical constants
\begin{equation*}
C_{1}=1,\quad C_{2}=1,\quad c_{0}=1,\quad d_{0}=1,\quad
b=-2\Longleftrightarrow \gamma =\frac{1}{2}
\end{equation*}
and
\begin{equation*}
\begin{array}{c|c|c|c}
n & m & \omega  & color \\ \hline
\frac{1}{2} & \frac{3}{2} & 1 & red \\ \hline
\frac{3}{2} & \frac{1}{2} & \frac{1}{3} & blue \\ \hline
\frac{1}{3} & \frac{1}{3} & 0 & magenta \\ \hline
\frac{-1}{3} & \frac{1}{2} & 1 & black
\end{array}
\end{equation*}
we have chosen the last values (black color) as pathological.

\begin{figure}[h!]
\begin{center}
\includegraphics[height=2.194in,width=2.194in]{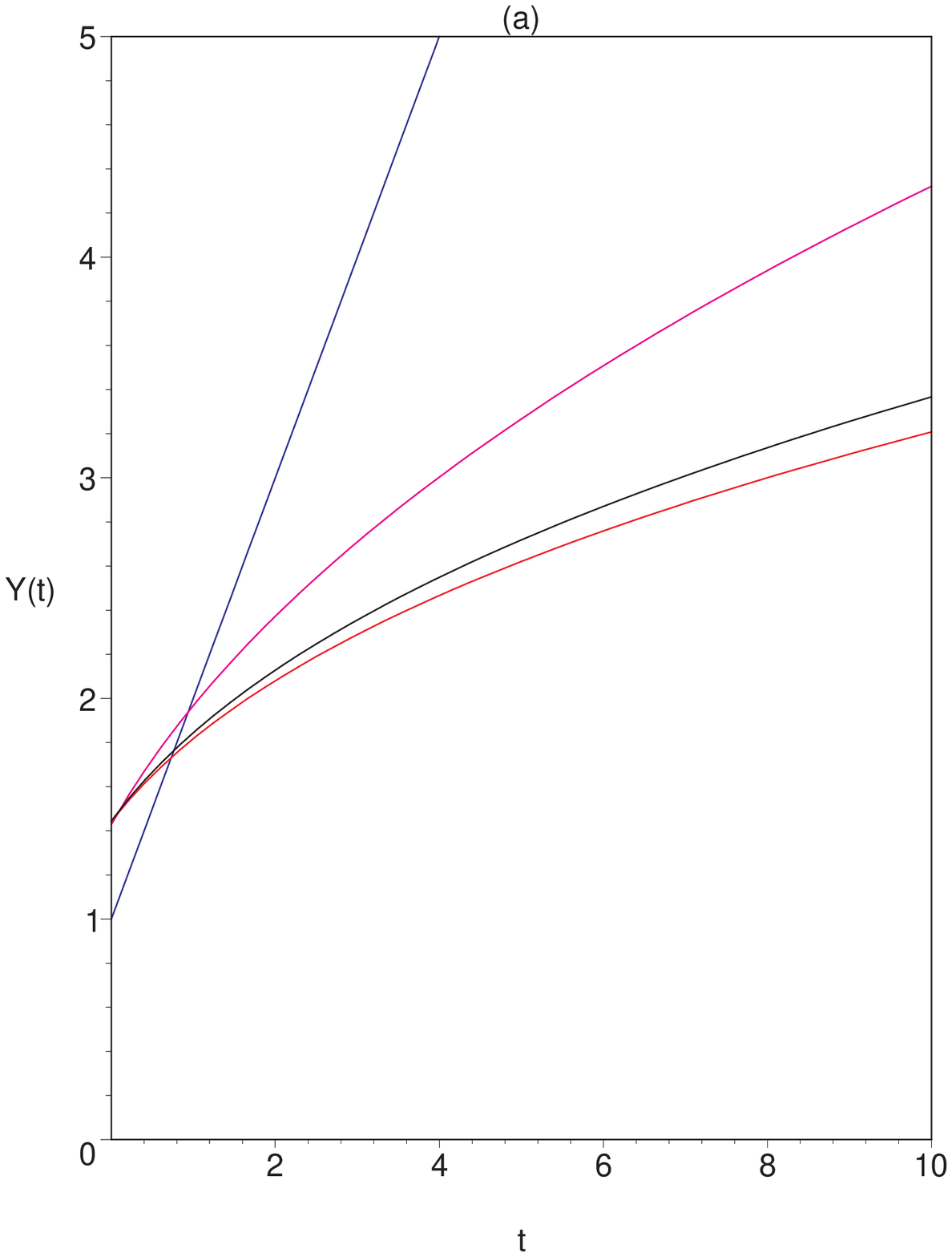} %
\includegraphics[height=2.194in,width=2.194in]{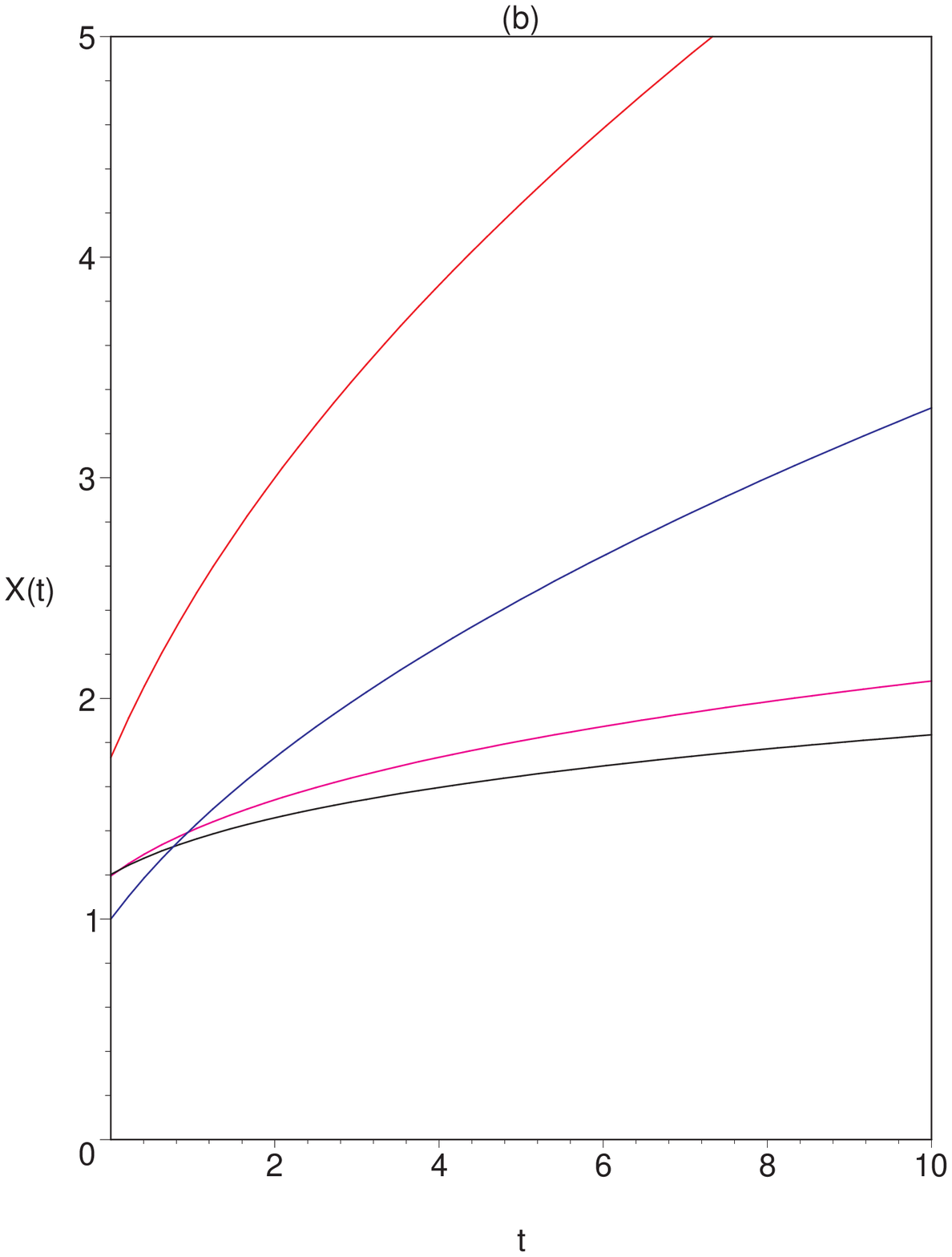}
\end{center}
\caption{In the figure (a) it is plotted the radius $Y(t)$ while in
figure (b) it is plotted the variation of the radius $X(t)$.
Both radius have a nonsingular start and are growing functions on time.}
\label{Radios3}
\end{figure}

With these numerical values we can see in fig. (\ref{Radios3}) the different
behaviour of our scale factors. In all the studied cases we can see that
these scale factors are growing functions on time $t$. These solutions are
non-singular as the Krestchmann invariants show us.

\begin{figure}[h!]
\begin{center}
\includegraphics[height=2.194in,width=2.194in]{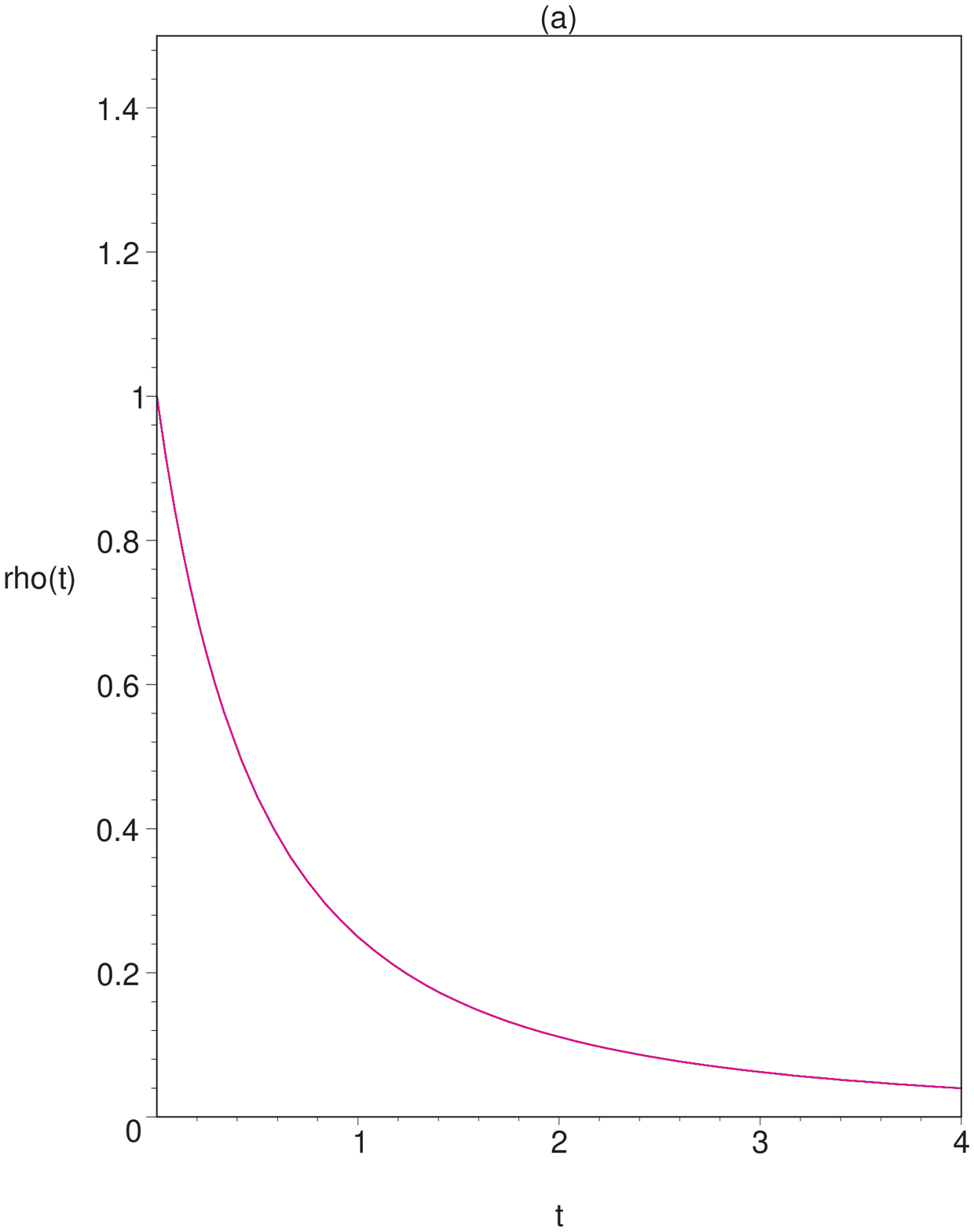} %
\includegraphics[height=2.194in,width=2.194in]{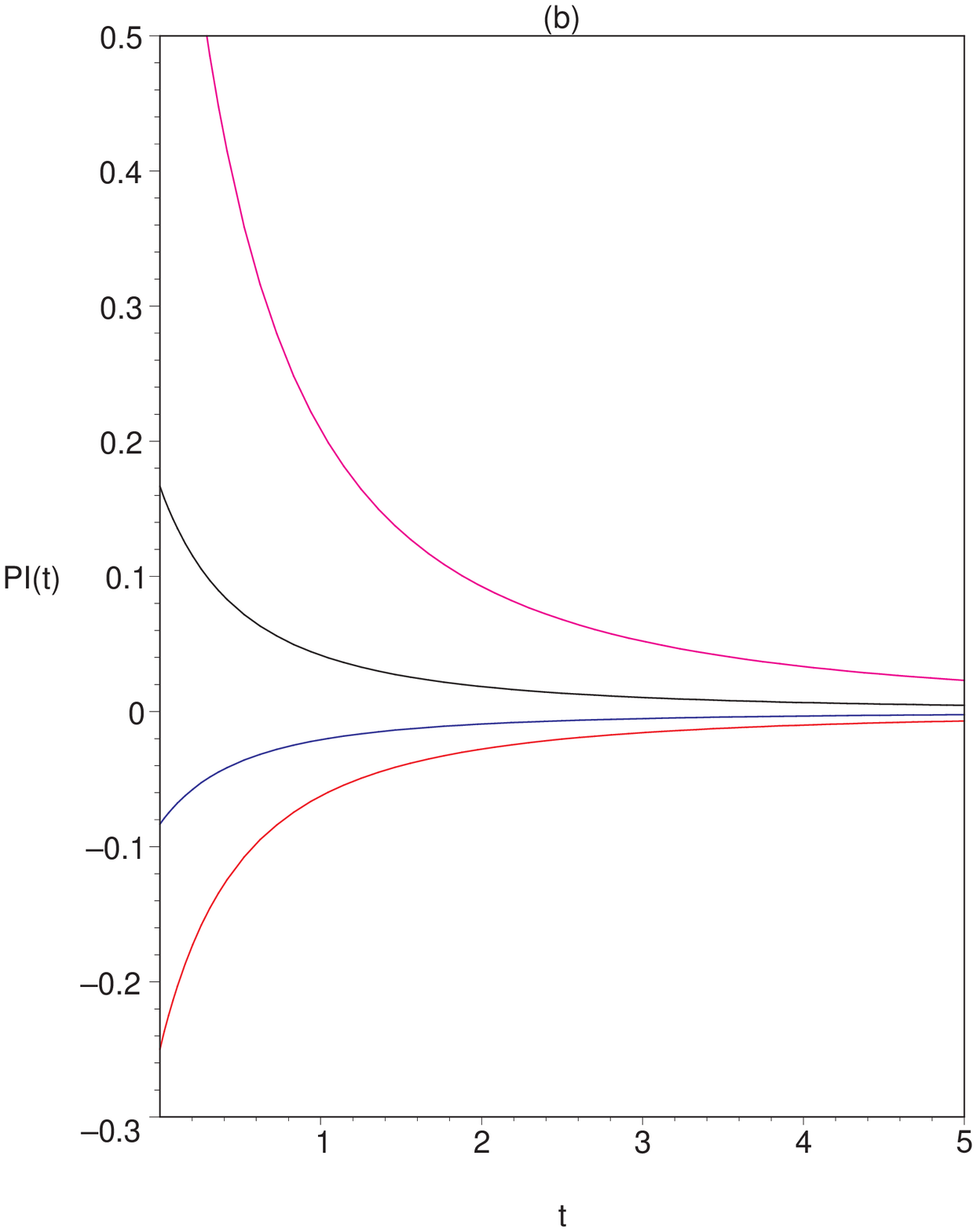}
\end{center}
\caption{Figure (a) shows the variation of the energy density $\protect%
\rho(t)$. Figure (b) shows the variation of the bulk viscous pressure $%
\Pi(t)$.}
\label{densidad3}
\end{figure}

With regard to the energy density and the viscous pressure (see fig. (\ref
{densidad3}))we can see that the only solutions with physical meaning are
those which have been plotted with red and blue colors i.e. with $n=\frac{1}{%
2},m=\frac{3}{2}$ and $\omega =1$ (ultrastiff matter) for the red color and $%
n=\frac{3}{2},m=\frac{1}{2}$ and $\omega =\frac{1}{3}$ for the
blue line. Since we have only been able to obtain a particular
solution (invariant solution) to the differential equation
(\ref{claudia}) for the energy density then this quantity shows
the same behaviour for the chosen numerical constants.

\begin{figure}[h!]
\begin{center}
\includegraphics[height=2.194in,width=2.194in]{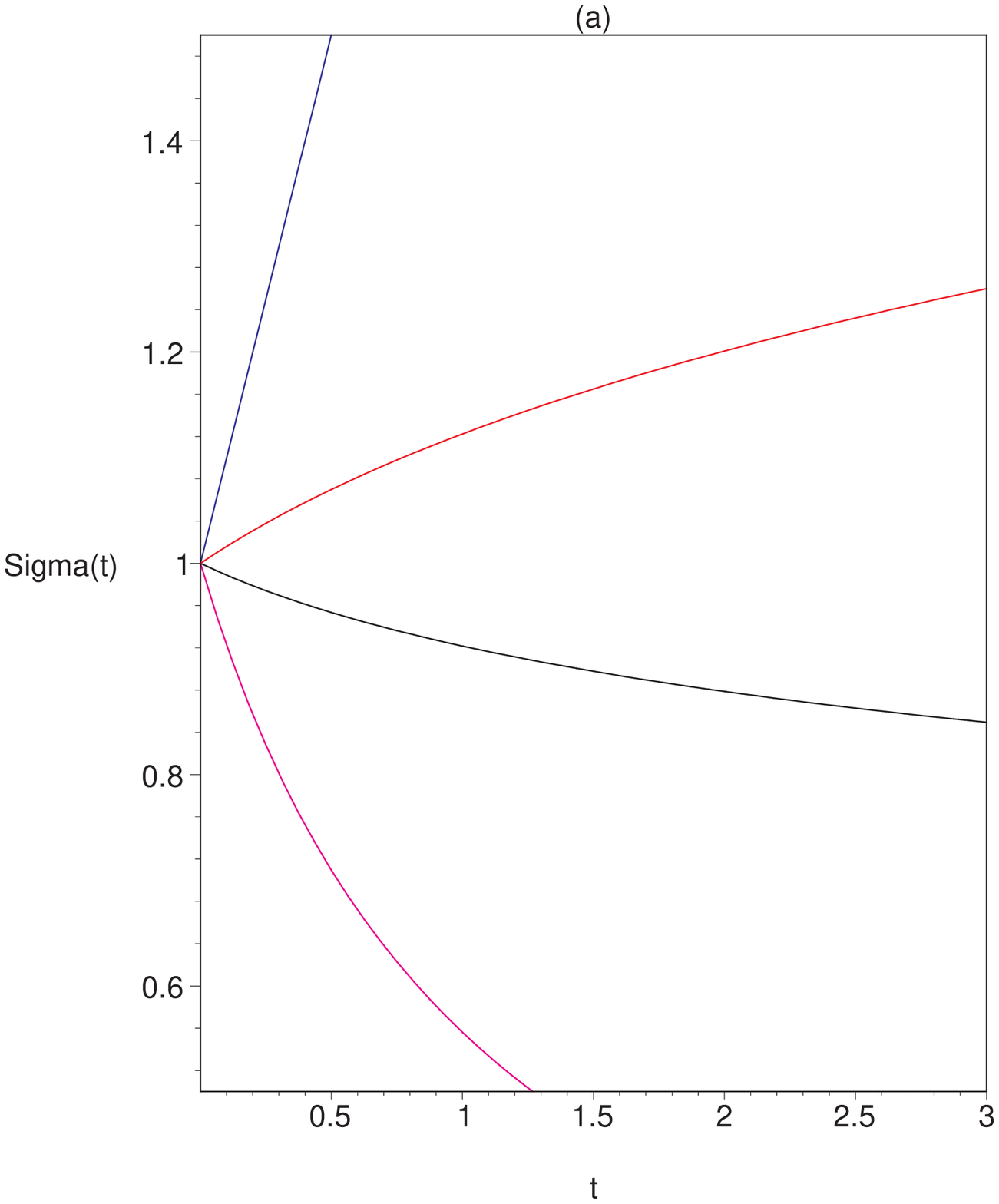} %
\includegraphics[height=2.194in,width=2.194in]{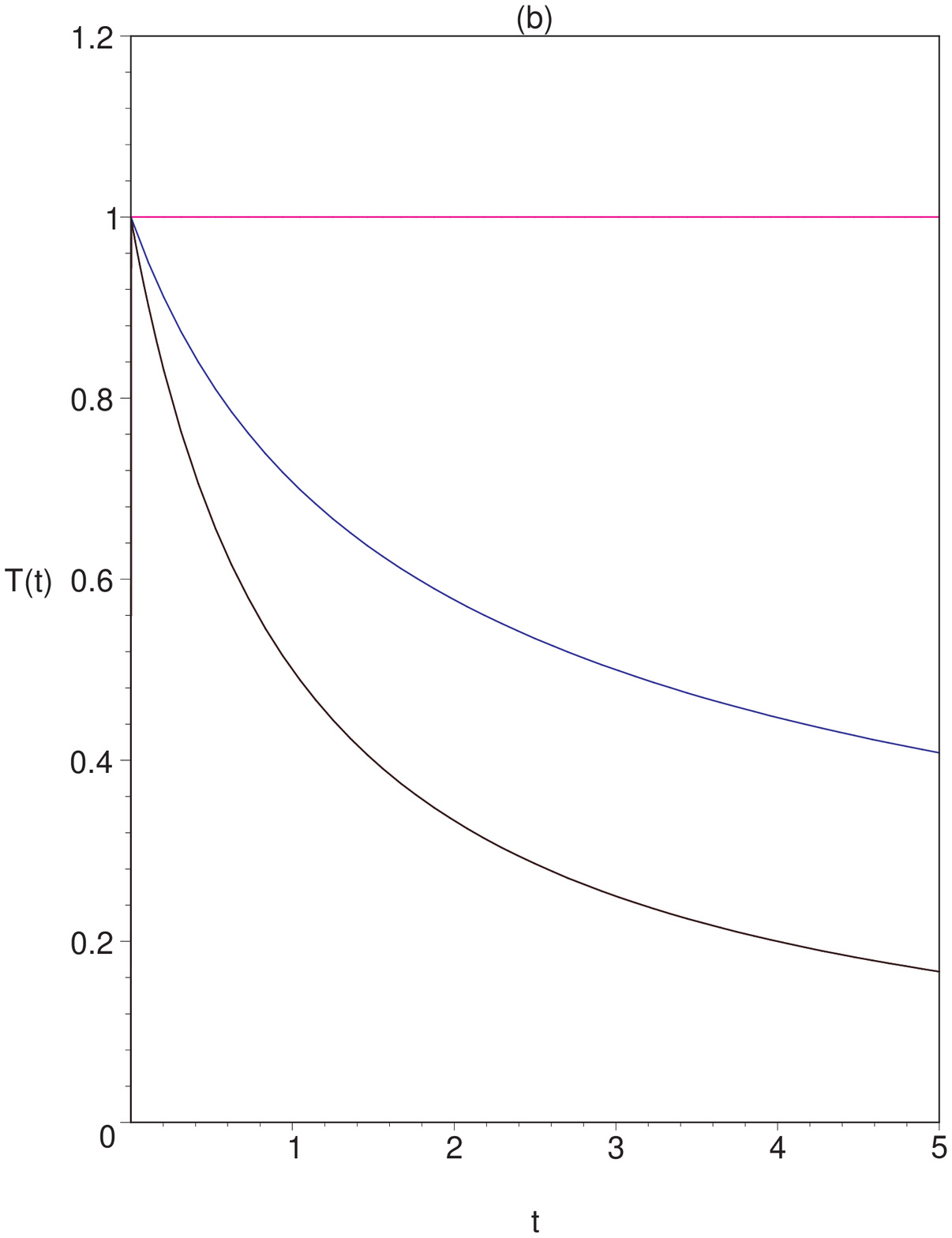}
\end{center}
\caption{Figure (a)  shows the variation of the entropy $\Sigma (t)$ while figure
 (b)  shows the variation of the temperature $T(t)$.}
\label{termo3}
\end{figure}

The behaviour of the thermodynamical quantities is showed in fig. (\ref
{termo3}). Even though all the studied cases shown a decreasing temperature
except the case $\omega =0$, magenta line, which correspond to $T(t)=const.,$
as it is expected. Picture (a) shows us that only the red and blue solutions
have physical sense as already we know from the last picture of the viscous
pressure the other solutions show a decreasing entropy $\Sigma (t)$.

\begin{figure}[h!]
\begin{center}
\includegraphics[height=2.194in,width=2.194in]{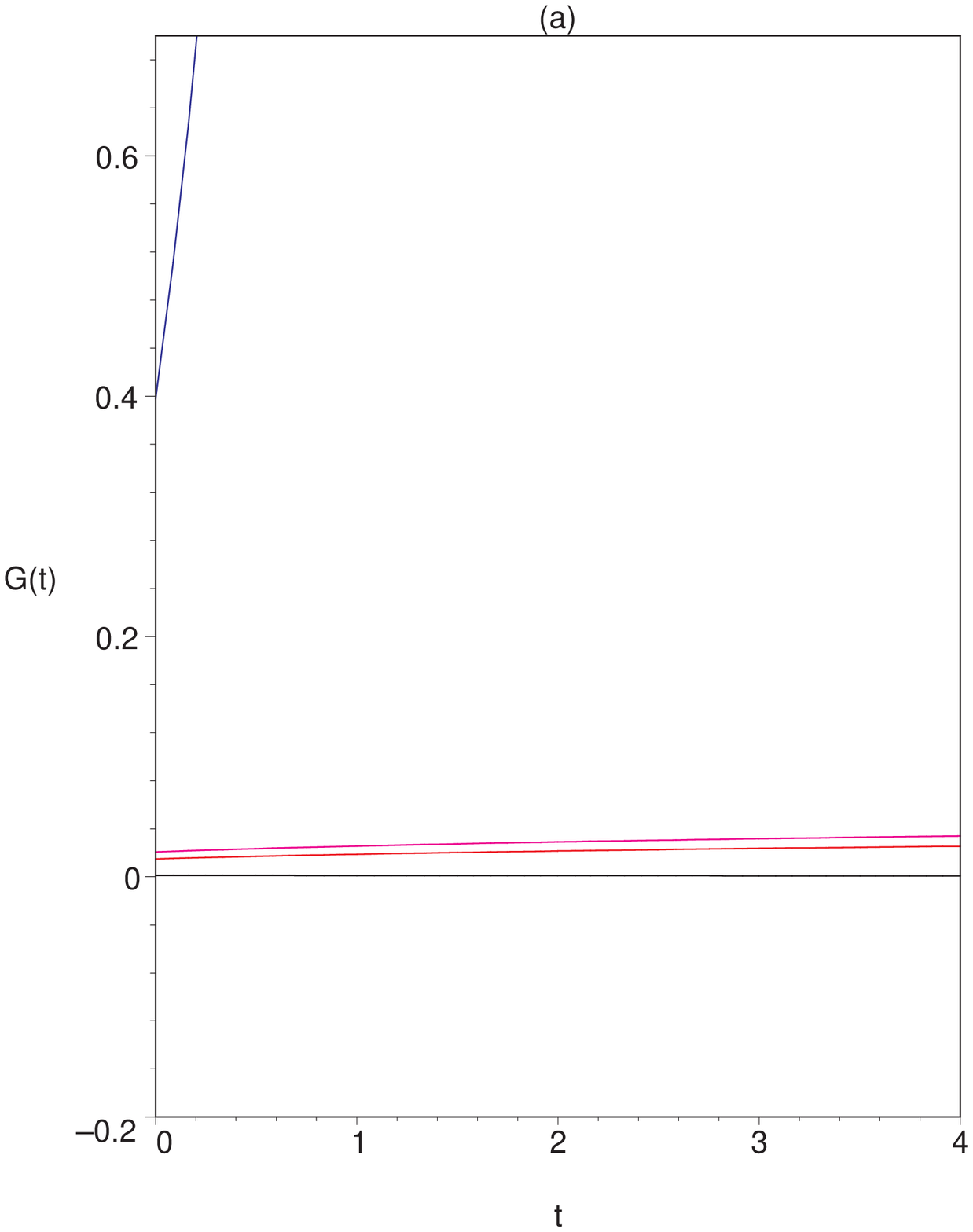} %
\includegraphics[height=2.194in,width=2.194in]{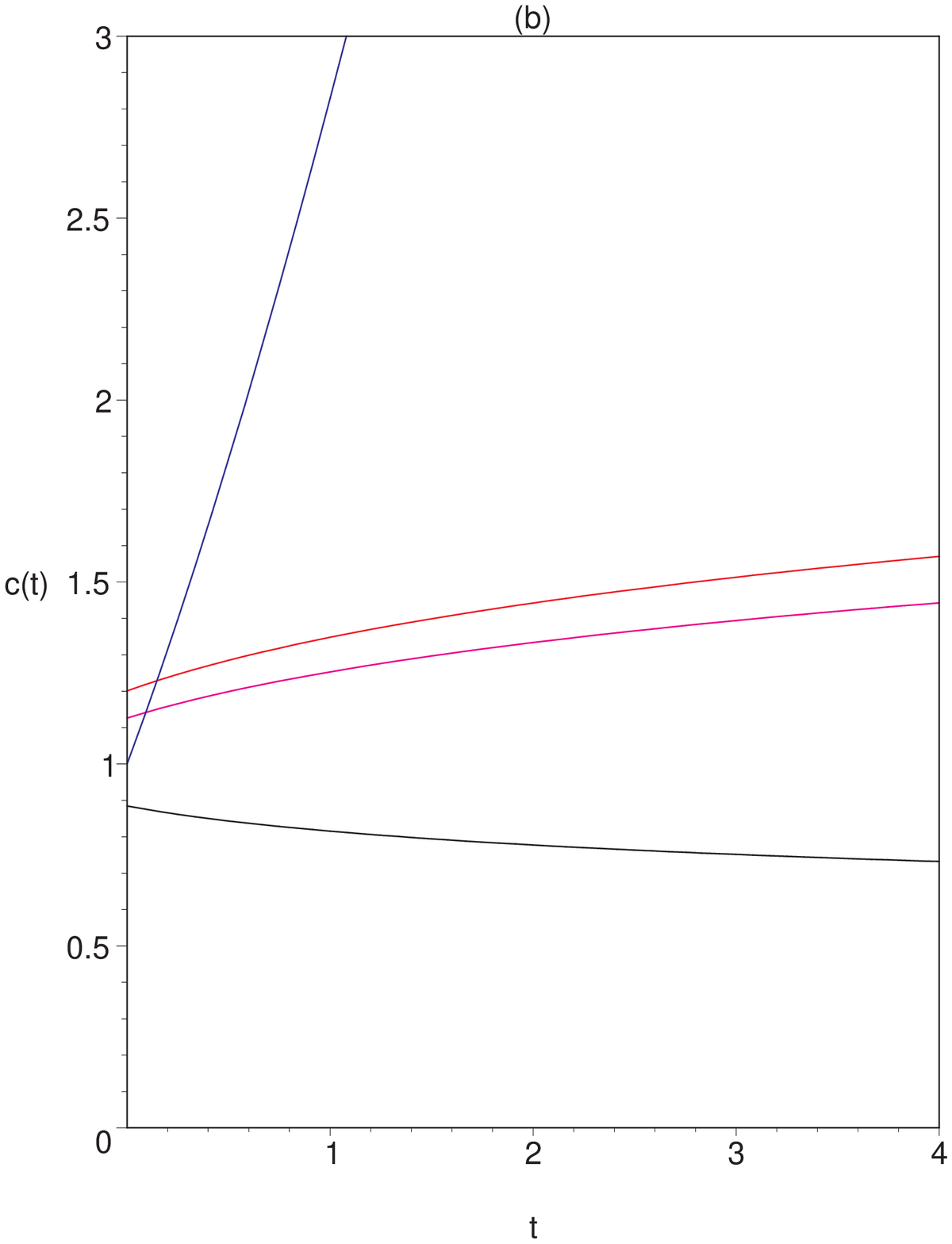} %
\includegraphics[height=2.194in,width=2.194in]{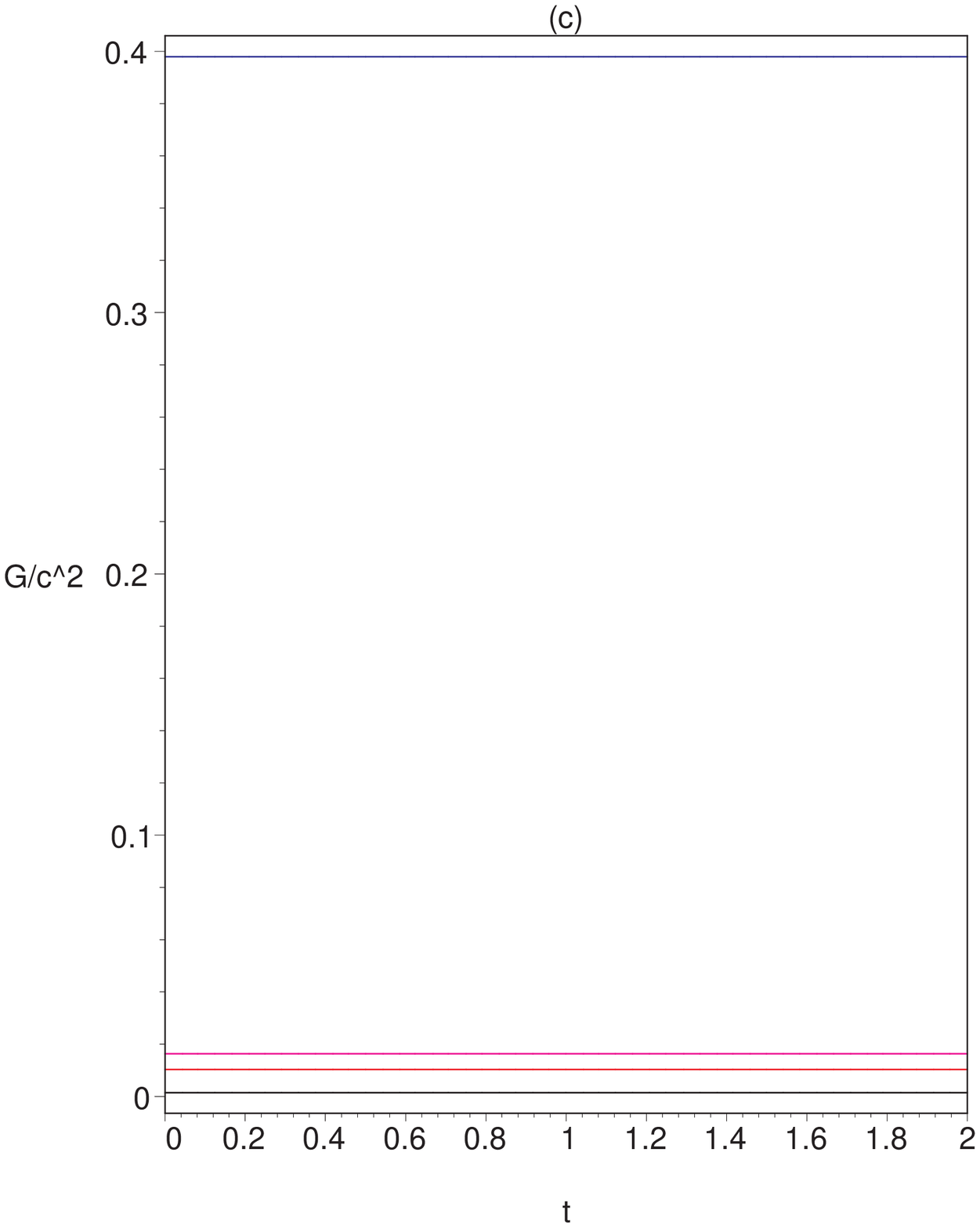}
\end{center}
\caption{In the figure (a) we show the variation of ``constant" $G(t)$. In the figure
 (b) we show the variation of the ``constant" $c(t)$ and in the figure (c) it is plotted the relationship
 $G/c^{2}$.}
\label{constantes3}
\end{figure}

The variation of the ``constants'' $G$ and $c$ as well as the relationship $%
G/c^{2}$ is shown in fig. (\ref{constantes3}). These pictures show
us that both ``constants'' are growing functions on time $t$
except in the case (black line) $n=-1/2$ which correspond to a
decreasing function. We would like to emphasize that only for the
case $\gamma =1/2$ we have that it is verified the relationship
$G/c^{2}=const.$

\begin{figure}[h!]
\begin{center}
\includegraphics[height=2.194in,width=2.194in]{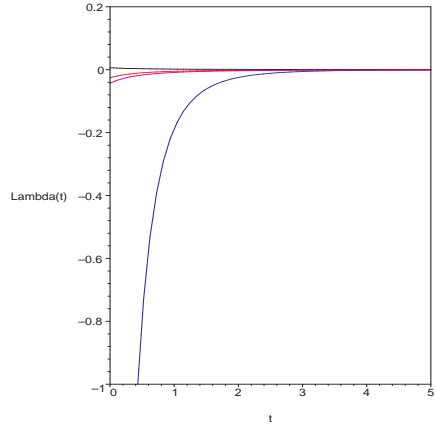}
\end{center}
\caption{The variation of the cosmological ``constant"  $\Lambda(t)$.}
\label{lambda3}
\end{figure}

With regard to the cosmological ``constant'', see fig.
(\ref{lambda3}), it is observed that all the solutions are
decreasing but negative except for the black line ($n=-1/2).$ In
these cases and because our models are non-singular $\Lambda
\longrightarrow \pm const$ when $t\longrightarrow 0.$

\begin{figure}[h!]
\begin{center}
\includegraphics[height=2.194in,width=2.194in]{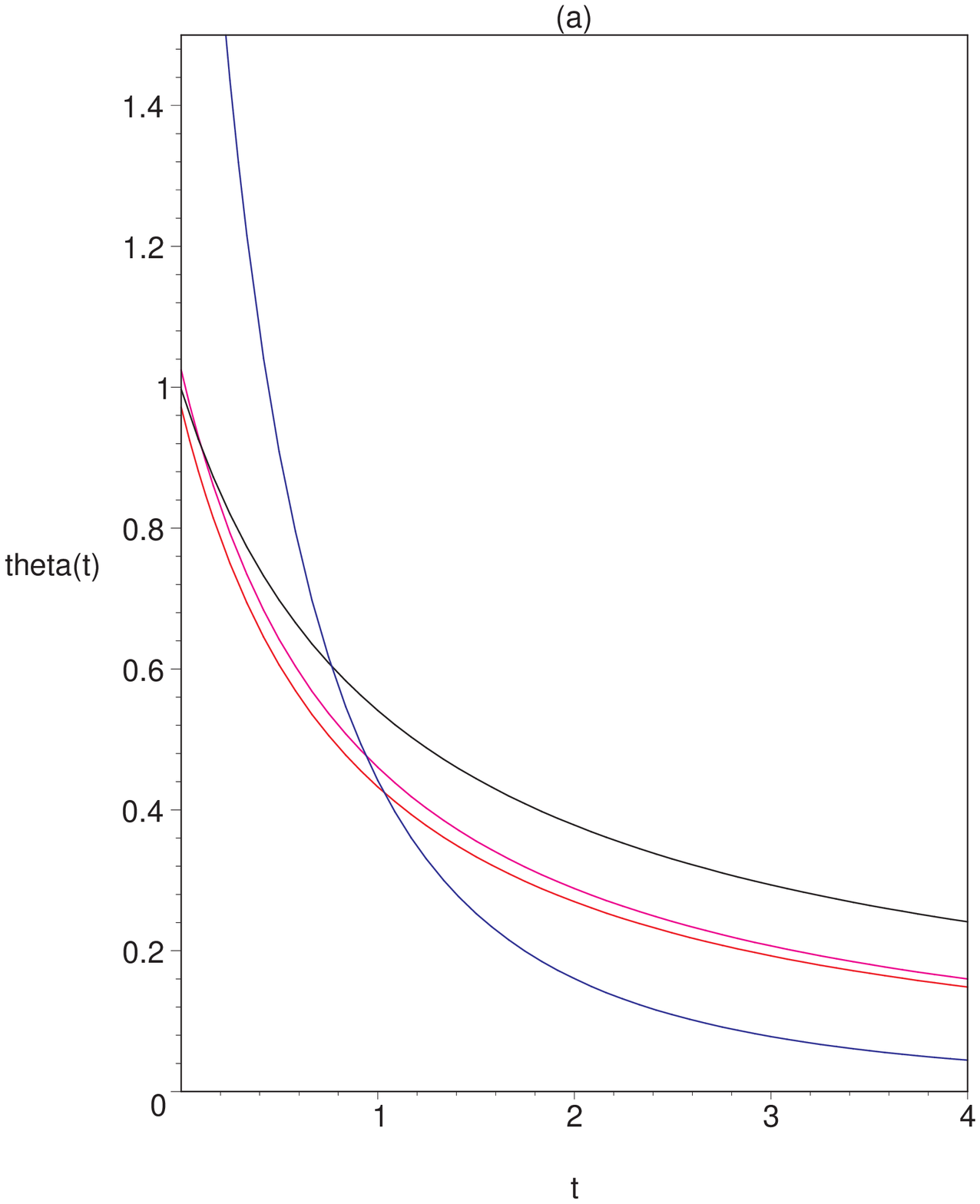} %
\includegraphics[height=2.194in,width=2.194in]{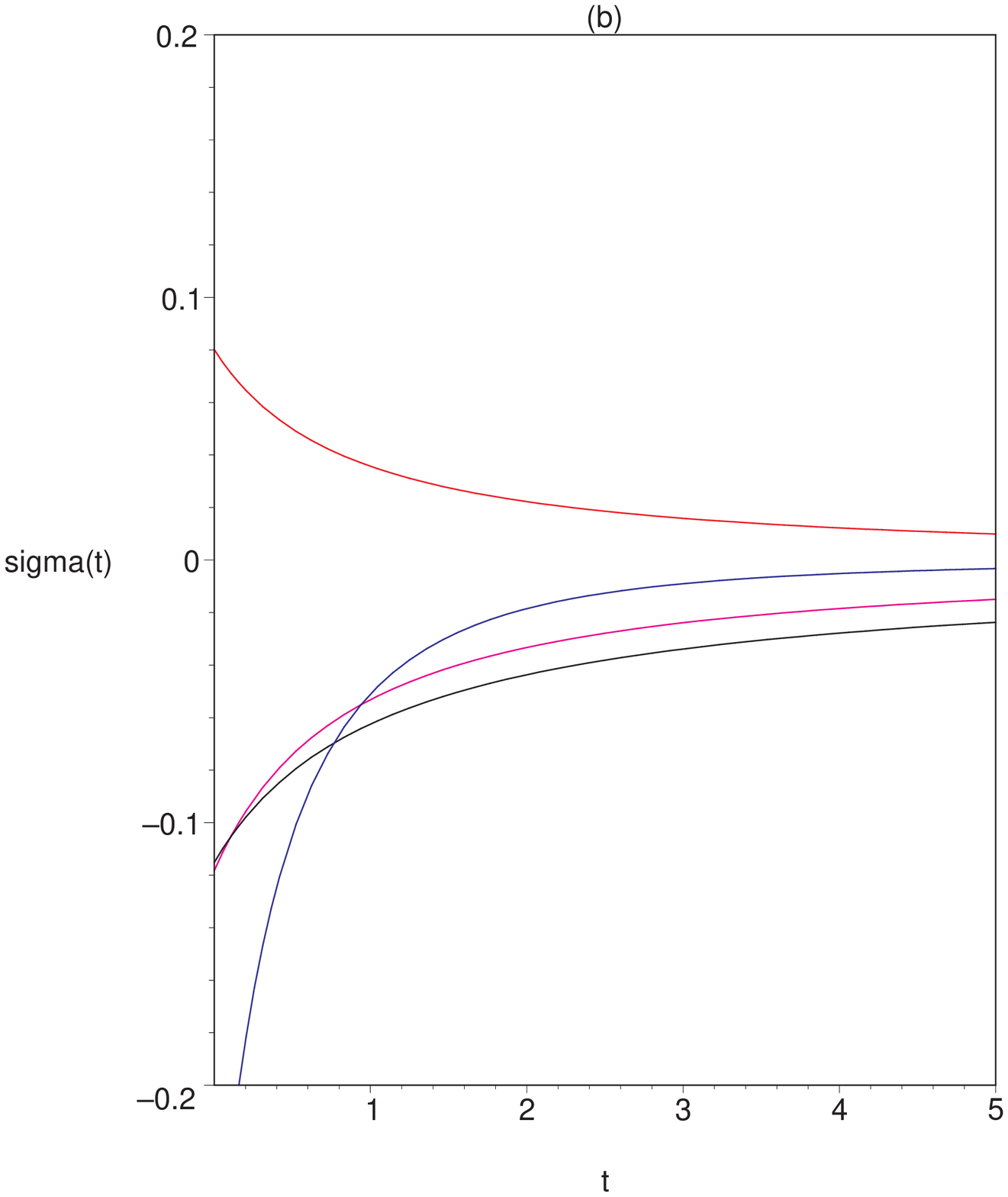}
\end{center}
\caption{In figure (a) it is plotted the expansion $\protect\theta (t)$,
while in figure (b) it is plotted the shear $\protect\sigma (t)$.}
\label{expansion3}
\end{figure}

The expansion and the shear behave as follows, see fig.
(\ref{expansion3}). As we can see all the models studied show a
decreasing expansion and only the models that have a positive
shear are plotted with the red and blue colors.

Other solutions can be studied changing the numerical values of the
constants as well as taking into account different equations of state. We
have found at least two non-singular physical solution (red and blue lines)
which equations of state are $\omega =1$ (ultrastiff \ matter) and $\omega
=1/3$ with $\gamma =1/2$ (the standard value for the viscous parameter).

\section{Conclusions and summary.}

In this paper we have studied three different cosmological
solutions for the presented model. In the first of them we have
needed to make two hypotheses in order to try to obtain a complete
solution of the field equations. The first assumption,
$X\thickapprox Y^{n}$, is standard in this class of models while
the second assumption, $\Pi \thickapprox \rho ,$formalizes through
the next equality, $\Pi =\varkappa \rho $, i.e. that the bulk
viscous pressure has the same order of magnitude of the energy
density, is more difficult of digesting. But as we have seen the
other two solutions we have founded precisely this solution. We
have to point out that, as we have seen, this solution is not the
most general solution for the outlined differential equation. The
relationship founded between the bulk viscous pressure and the
energy density is the invariant solution, a particular solution.
Maybe the most general solution to this ode brings us to obtain
other relationship between these quantities but as it has been
showed in (\cite{TonyCas}) for the classic FRW model the general
solution to the outlined differential equation in the latter (see
the appendix of (\cite{TonyCas}))  has no physical meaning while
the invariant solution (the classical solution) at least has
physical sense. For all these reason together with the reasons
exposed above, we believe that this hypothesis is correct.

For the first of our models we find that this is singular and
thermodynamically correct since the entropy is a growing function on time $t$%
, if and only if $\varkappa \neq 0,$ while the temperature
decreases. With regard to the behaviour of the ``constants'', it
is founded that the
constants $G$ and $c$ for $\gamma =1/2$ must verify the relationship $%
G/c^{2}=const.$ in spite of the fact that both constants vary but
in such a way that this quotient remains constant for all $t.$

With regard to the second of our models we can see that the red color
solution which corresponds to $\omega =1$ and $\gamma =1/2$ is the only
physical solution. This solution has been obtained under the assumptions
that $X\thickapprox Y^{n}$ and $c\thickapprox t^{a}$, i.e. the ``constant'' $%
c(t)$, the speed of light, follows an power law dependence of time
$t.$ Both hypotheses seem reasonable. The red solution is singular
and both ``constants'' $G$ and $c$ are growing functions on time
$t$ under the imposed restrictions in order to find a growing
entropy $\Sigma $ and a bulk viscous pressure verifying the
condition $\Pi <0.$ As we have commented above for this case it is
founded that $\Pi \thickapprox \rho $, i.e. both quantities have
the same order of magnitude. For the studied viscous
parameter $\gamma =1/2$, and only for it, we have showed that relationship $%
G/c^{2}=const.$ is verified for both constants.\ The cosmological constant
is a negative decreasing function on time $t.$

The third of our solutions is non-singular and have been obtained
under the assumptions $c\thickapprox Y^{n}$ and $X\thickapprox
Y^{m}.$ In this occasion we have obtained two physical solutions
which correspond to the red and blue colors, with  $\omega =1$ and
$\omega =1/3$ respectively and in both cases the viscous parameter
is $\gamma =1/2,$ the usual one. The behavior obtained for the
main quantities is similar to the obtained in for the second of
our models except that in this case the solutions are non-singular
i.e. the constants have a growing behavior while the cosmological
constant is a negative decreasing function.

\vspace{0.5cm}
\textbf{Acknowledgement} I wish to acknowledge to Javier Aceves his translation into
English of this paper.

\end{document}